\documentclass[twocolumn,superscriptaddress,longbibliography,aps,pra,preprintnumbers,10pt]{revtex4-2}

\usepackage[utf8]{inputenc}
\usepackage[urlcolor=blue,colorlinks=true,citecolor=blue,linkcolor=blue,pdfstartview={FitH},bookmarks=false]{hyperref}
\usepackage[T1]{fontenc}
\usepackage{graphicx}
\usepackage{xcolor}
\usepackage{amsmath,amssymb,amsfonts}
\usepackage{verbatim}
\usepackage{ulem}

\newcommand{\JB}[1]{\textcolor{black}{#1}}

\newcommand{\wymiarm}{0.4\textwidth}

\begin{document}

\preprint{}

\title{Interplay of correlations and Majorana mode from local solution perspective}

\author{Jan Bara\'nski}
\email[e-mail: ]{j.baranski@law.mil.pl (Corresponding author)}
\affiliation{Department of General Education, Polish Air Force University, ul. Dywizjonu 303 nr 35, 08521 D\k{e}blin, Poland}

\author{Magdalena Bara\'nska}
\email[e-mail: ]{m.baranska@law.mil.pl}
\affiliation{Department of General Education, Polish Air Force University, ul. Dywizjonu 303 nr 35, 08521 D\k{e}blin, Poland}

\author{Tomasz Zienkiewicz}
\email[e-mail: ]{t.zienkiewicz@law.mil.pl}
\affiliation{Department of General Education, Polish Air Force University, ul. Dywizjonu 303 nr 35, 08521 D\k{e}blin, Poland}

\author{Tadeusz Doma\'nski}
\email[e-mail: ]{doman@kft.umcs.lublin.pl}
\affiliation{\mbox{Institute of Physics, M. Curie-Sk\l{}odowska University, 20-031 Lublin, Poland}}

\date{\today}

\begin{abstract}
We study the quasiparticle spectrum of a hybrid system, comprising \JB{a} correlated (Anderson-type) quantum dot coupled to \JB{a} topological superconducting nanowire hosting the Majorana boundary modes. From the exact solution of the \JB{low-energy} effective Hamiltonian, we uncover \JB{a} subtle interplay between Coulomb repulsion and the Majorana mode. Our analytical expressions show that the spectral weight of the leaking Majorana mode is sensitive to \JB{both} the quantum dot energy level and the repulsive potential. We \JB{compare} our results with estimations by L.S. Ricco {\it et al.} \href{https://link.aps.org/doi/10.1103/PhysRevB.99.155159}{Phys. Rev. B {\bf 99}, 155159 (2019)} obtained for the same hybrid structure \JB{using} the Hubbard-type decoupling scheme, and analytically quantify the spectral weight of the zero-energy (topological) mode coexisting with the finite-energy (trivial) states of \JB{the} quantum dot. We also show that empirical verification of these spectral weights could be feasible \JB{through} spin-polarized Andreev spectroscopy.
\end{abstract}

\maketitle

\section{Introduction}
Quantum dots side-attached to topological superconducting nanowires have been considered as a \JB{suitable} platform for probing the Majorana boundary modes \cite{Baranger-2011,Deng-2016,Li2015,Liu2015,Gong2014,Leijnse2014,Ricco-2019}
which can demonstrate their non-local nature \cite{Prada2017}.
Hybridization between these constituents induces the intersite pairing, allowing for leakage of the Majorana mode onto the quantum dot region. Such \JB{a} process has been initially predicted for the uncorrelated case \cite{Vernek-2014} and later on also in \JB{the presence} of Coulomb repulsion \cite{Lopez-2013,Lutchyn-2014,Moca-2014,Maciejko-2019,Silva-2020,Orellana2020,Orellana2023,Zienkiewicz2016}.
Distinguishing the Majorana zero modes (MZMs) from trivial states of \JB{the} QD is, however, a challenging issue because various trivial states at zero energy could mimic the behavior of MZMs \cite{Prada2017,Valentini2022}. 

Furthermore, in various hybrid \JB{structures} the trivial states can coexist with topological ones \cite{Chen2019,Hess2021} and their signatures might potentially yield misleading conclusions. For example, Liu {\it et al.} \cite{Liu2017} demonstrated that coalescence of the Andreev states can enhance zero-bias conductance to $2e^2/h$, typical for the Majorana mode. Kondo resonance, appearing at zero energy in strongly correlated structures, could \JB{also} be confused with the Majorana quasiparticle. Differences between these effects could be resolved by spin-polarized tunneling spectroscopy \cite{Lopez-2013,Lutchyn-2014,Moca-2014,Maciejko-2019,Silva-2020,Golub2011,Wojcik2017,Majek2024}, yet their unambiguous identification would be rather difficult.

\begin{figure}
	\includegraphics[width=\wymiarm]{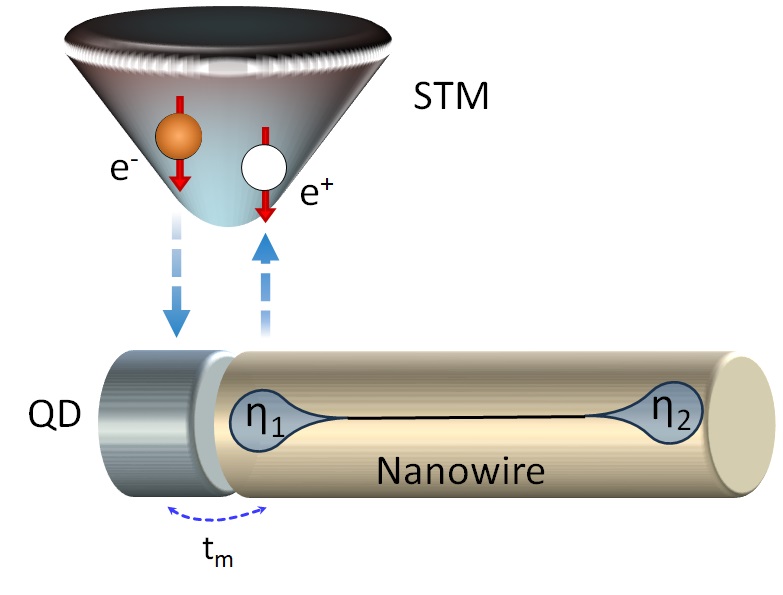}
	\caption{Schematics of the quantum dot (QD) attached to the topological nanowire, hosting the boundary Majorana modes $\eta_{i}$. Quasiparticles of \JB{the} QD could be probed by spin-polarized scanning spectroscopy, measuring the conductance of the charge current contributed by electron-to-hole (Andreev) scattering of identical spins (marked by red arrows).
} 
	\label{fig.schem}
\end{figure}

Given these facts, there is \JB{an} ultimate need to accurately describe the quasiparticle spectra in topological hybrid systems. To address this issue, we analyze here the minimal setup composed of the Anderson-type quantum impurity coupled to the Majorana mode (Fig. \ref{fig.schem}), which can be solved analytically. From the exact solution, we determine the eigenstates and analytically express the quasiparticle energies and their spectral weights, providing information about optimal conditions for leakage of the zero-energy Majorana mode onto the correlated quantum impurity with \JB{strong} Coulomb repulsion between opposite-spin electrons. Our study could be regarded as complementary to the previous investigations based either on the Hubbard-I decoupling scheme \cite{Ricco-2019} or other purely numerical considerations \cite{Lopez-2013,Lutchyn-2014,Moca-2014,Maciejko-2019,Silva-2020,Orellana2020,Orellana2023}. Information derived from such analytical results could be useful for considerations of these quantum dot-topological superconductor hybrid structures under nonequilibrium conditions (for instance imposed by gate potentials or time-dependent driving) when precise knowledge of the eigenfunctions and quasiparticle energies is necessary to deduce the quantum evolution.

For experimental detection of the Majorana and the trivial bound states, we consider the Selective Equal-Spin Andreev Reflection (SESAR) spectroscopy.  
In contrast to ordinary Andreev reflection, its mechanism relies on polarized charge transfer by scattering an electron into a hole of the same spins. In quantum dot-Majorana hybrids, this process is feasible due to the intersite triplet pairing. Such a mechanism was proposed by He {\it et al.} \cite{He2014} for reliable identification of Majorana quasiparticles. Spin-polarized Andreev spectroscopy has also enabled the detection of topological zero-energy modes inside the vortex in a topological superconductor \cite{Sun2016}. 
In subsequent studies, spin-dependent transport characteristics have been measured for magnetic atom chains, 
revealing inherent polarization of the Majorana quasiparticles at their edges \cite{Zhang2018}. 
SESAR has also been proposed for probing the spatial profile of Majorana quasiparticles in topological planar Josephson junctions \cite{Godzik2020}.

Furthermore, the recent realization of \JB{the} minimal Kitaev chain in \JB{double} quantum dots interconnected through \JB{a} conventional superconductor \cite{Dvir2023,Souto2023} enabled \JB{the} realization of triplet pairing, which has been resolved by \JB{spin-polarized} crossed Andreev scattering \cite{Wang2022}. Another platform for \JB{Majorana} quasiparticles \JB{are topological} nodal-point superconductors \cite{Zeng2024}, where SESAR spectroscopy has been used as well.
Motivated by \JB{the} popular use of spin-resolved Andreev spectroscopy, we inspect its mechanism here in the minimal QD-MBS setup, providing the exact Green's functions, which \JB{encode} information on the SESAR processes.

It has been established \cite{Leijnse2011} that charge tunneling could probe the lifetime of the Majorana states in heterostructures consisting of \JB{a} metal-quantum dot-topological superconductor. Charge transfer varies the electron number on the quantum dot by $\pm 1$, thus connecting the even \JB{and} odd parity sections. In what follows, we determine \JB{the} probability of such parity changes in the strongly correlated quantum dot. This brings information \JB{concerning} optimal conditions for \JB{the} leakage of the Majorana modes.

The paper is organized as follows. In Sec.\ \ref{Sec.two} we introduce the model and present \JB{the} general forms of its \JB{eigenstates and eigenenergies} for arbitrary overlap between the Majorana boundary modes. Next, in Sec.\ \ref{Sec.three}, we analyze the spin-resolved quasiparticle spectra of the correlated QD coupled only to one Majorana mode. The next Sec.\ \ref{Sec.four} generalizes our treatment \JB{to} the case with nonzero overlap between the Majorana modes. Finally, we summarize the obtained results. \JB{The} Appendix provides brief information \JB{concerning} the role of \JB{the} magnetic field.

\section{Eigenstates and eigenenergies}
\label{Sec.two}

The low-energy physics of the hybrid structure shown in Fig. \ref{fig.schem} can be described by the following Hamiltonian:
\begin{eqnarray}
    \hat{H}=\hat{H}_{QD}+\lambda(\hat{d}_{\downarrow}^{\dagger} \hat{\eta}_{1}+\hat{\eta}_{1}\hat{d}_{\downarrow})+i\epsilon_{m} \hat{\eta}_{1}\hat{\eta}_{2} ,
    \label{H0}
\end{eqnarray}
where
\begin{eqnarray}
\hat{H}_{QD}=\sum_{\sigma} \varepsilon_{d} \hat{d}_{\sigma}^{\dagger}\hat{d}_{\sigma}+U_d \hat{n}_{\uparrow}\hat{n}_{\downarrow}
\end{eqnarray}
refers to the correlated quantum dot (QD) with the energy level $\varepsilon_{d}$ and the Coulomb potential $U_{d}$. The second term on \JB{the} r.h.s. of Eq.~(\ref{H0}) describes the coupling of \JB{the} QD to one of the boundary states, $\hat{\eta}_{1}$, of \JB{the} topological nanowire. In the analyzed model, we assume that the tunneling between the dot and the MZM is spin-polarized. This is because Majorana modes in topological superconductors are typically associated with a specific spin polarization, depending on the direction of the magnetic field and the spin-orbit interaction. The boundary modes are described by self-hermitian operators $\hat{\eta}_{i}^{\dagger}=\hat{\eta}_{i}$. The last term stands for an overlap between the Majorana modes ($\hat{\eta}_{1}$, $\hat{\eta}_{2}$) and it is relevant to short topological nanowires.

It is convenient to express the Majorana operators in terms of the conventional fermion operators $\hat{f},\hat{f^{\dagger}}$ defined through $\hat{\eta}_{1}=\frac{1}{\sqrt{2}}(\hat{f}^{\dagger}+\hat{f})$ and $\hat{\eta}_{2}=\frac{i}{\sqrt{2}}(\hat{f}^{\dagger}-\hat{f})$. Hamiltonian (\ref{H0}) \JB{then acquires} the following structure:

\begin{eqnarray}
    \hat{H}&=&\hat{H}_{QD}+t_m(\hat{d}_{\downarrow}^{\dagger}\hat{f} +\hat{f}^{\dagger}\hat{d}_{\downarrow}) \nonumber \\
    &+&t_m(\hat{d}_{\downarrow}^{\dagger}\hat{f}^{\dagger}+fd_{\downarrow})+ \epsilon_{m}(\hat{f}^{\dagger}\hat{f}-\frac{1}{2}) ,
\label{model_Hamil}    
\end{eqnarray}
where $t_{m}=\lambda/\sqrt{2}$.
We note that the second part of this Hamiltonian (\ref{model_Hamil}) represents the usual tunneling of a spin-$\downarrow$ electron between the QD and the topological nanowire, while the third part represents the intersite pairing potential, where triplet pairs are formed or annihilated. 

\begin{figure}[b!]
	\includegraphics[width=\wymiarm]{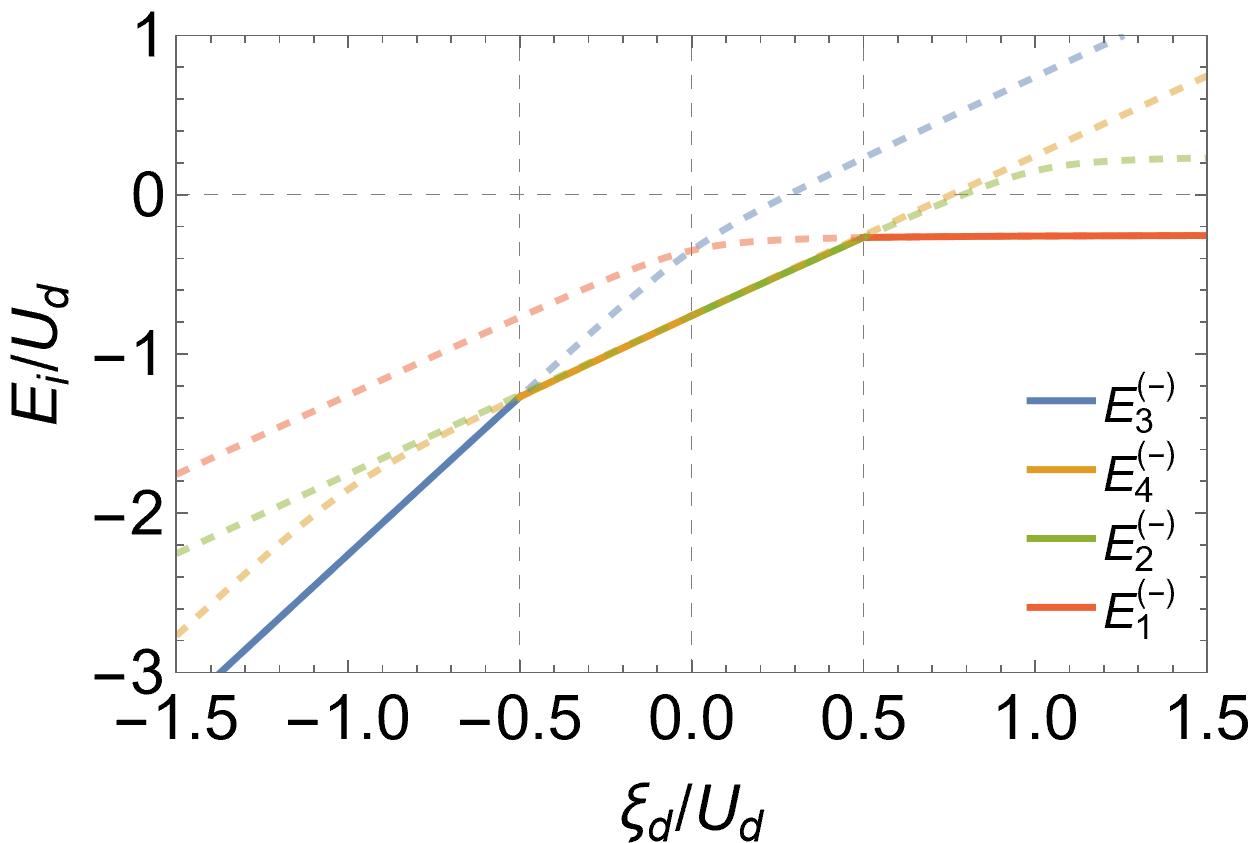}
	\caption{Dependence of the eigenenergies $E_{i}^{-}$ on the energy level $\varepsilon_{d}$ of \JB{the} QD. Solid lines refer to the ground-state energy. Results are obtained for $t_m=0.1U_d$ and $\epsilon_m=0.5U_d$. Dashed faded lines represent excited states.
} 
	\label{fig.grE}
\end{figure}

\JB{The} Hilbert space of the model Hamiltonian (\ref{model_Hamil}) is spanned by eight states $|n_{d\sigma},n_{f}\rangle$. Its eigenstates can be determined analytically and are represented by the following superpositions:
\begin{eqnarray}
    |\Psi^{\pm}_{1} \rangle &=& u^{\pm}_{1}|0,0\rangle+v_1^{\pm}|\downarrow,1\rangle ,
    \label{eq4} \\ 
    |\Psi_{2}^{\pm} \rangle &=&u_{2}^{\pm}|\downarrow,0\rangle+v_{2}^{\pm} |0,1\rangle ,
    \label{eq5} \\ 
    |\Psi_{3}^{\pm} \rangle &=& u_{3}^{\pm} |\uparrow \downarrow,0\rangle +v_{3}^{\pm}|\uparrow,1\rangle ,
    \label{eq6} \\ 
    |\Psi_{4}^{\pm}\rangle&=& u_{4}^{\pm} |\uparrow,0\rangle+v_{4}^{\pm} |\uparrow\downarrow,1\rangle .
    \label{eq7}
\end{eqnarray}
Let us remark \JB{ that} the correlated quantum dot coupled to \JB{a} conventional superconductor would be characterized by a different set of eigenvectors, represented either by the singly occupied configurations $|\uparrow\rangle$ and $|\downarrow\rangle$ or the BCS-type coherent superpositions $u^{\pm} |0\rangle+v^{\pm}|\uparrow\downarrow\rangle$ \cite{Bauer2007,Baraski2013}.
Here, in contrast, we obtain eigenstates that are superpositions of either the empty and singly occupied dot, $|\Psi_{1,2}\rangle$, or the doubly and singly occupied dot, $|\Psi_{3,4}\rangle$, combined with the edge mode. Unlike the mentioned BCS-type superpositions, the eigenstates of the considered system are superpositions of states with different dot electron parity.
One can also note that the states $|\Psi_{1,2}\rangle$ are characterized by opposite dot magnetization compared to states $|\Psi_{3,4}\rangle$. Consequently, a ground state transition from $|\Psi_{1,2}\rangle$ to $|\Psi_{3,4}\rangle$ (or vice versa) is accompanied by \JB{a} conversion of \JB{the dot's} magnetic properties (c.f. Fig. \ref{fig.grE}). 
Such a set of eigenfunctions originates from the intersite pairing. For each of these configurations, we obtained two possible solutions, $\hat{H}|\Psi_{i}^{\pm}\rangle=E_{i}^{\pm}|\Psi_{i}^{\pm}\rangle$, with eigenvalues
\begin{eqnarray}
    E_{1}^{\pm}&=&\frac{1}{2}\left[ \epsilon_{d} \pm \sqrt{(\epsilon_{d}+\epsilon_{m})^2+4t_m^2} \right] ,
    \label{eq8a} \\ 
     E_{2}^{\pm}&=&\frac{1}{2}\left[ \epsilon_{d} \pm \sqrt{(\epsilon_{d}-\epsilon_{m})^2+4t_m^2} \right] ,
     \label{eq9} \\ 
     E_{3}^{\pm}&=&\frac{1}{2}\left[ 3\epsilon_{d} +U_d \pm \sqrt{(\epsilon_{d}-\epsilon_{m}+U_{d})^2+4t_m^2} \right] ,
     \label{eq9b} \\ 
     E_{4}^{\pm}&=&\frac{1}{2}\left[ 3\epsilon_{d} +U_d \pm \sqrt{(\epsilon_{d}+\epsilon_{m}+U_{d})^2+4t_m^2} \right] .
     \label{eq10}
\end{eqnarray}
As the values of the square roots are positive\JB{,} therefore candidates for the ground state are only those eigenenergies (\ref{eq8a}-\ref{eq10}) with \JB{a} minus-sign in front of the square root.

Figure \ref{fig.grE} illustrates these eigenenergies, indicating the ground state energy (solid line) obtained for large overlap between \JB{the} QD and the Majorana modes, $\epsilon_{m}=0.5U_{d}$. The dashed faded lines mark the eigenvalues of the excited states. The eigenenergies are plotted with respect to the parameter $\xi_d = \epsilon_d + \frac{U_d}{2}$, which represents \JB{the} departure from the half-filled QD.
In Fig.\ \ref{fig.grE} we used the on-site Coulomb repulsion as the energy unit in order to highlight the critical points at which \JB{the} ground state is represented by different types of states.
Note that configurations $|\Psi_{3}\rangle$ and $|\Psi_{4} \rangle$ have a component related to the double QD occupancy, while for the states $|\Psi_{1}\rangle$ and $|\Psi_{2}\rangle$ a maximum number of electrons on \JB{the} QD is one. Therefore, below QD half-filling $(\xi_d<0)$ the ground state is represented either by $|\Psi_{3}\rangle$ or $|\Psi_{4} \rangle$. In the opposite case, the Coulomb potential imposes the ground state $|\Psi_{1}\rangle$ or $|\Psi_{2}\rangle$. 

In what follows, we shall inspect the quasiparticle excitation spectrum
that could be probed by tunneling experiments when our hybrid structure is contacted with a conducting tip. \JB{The main} purpose of this study is to evaluate the spectral weights shared between the topological and trivial branches, upon varying the energy level of \JB{the} correlated QD.

\section{Results for $\epsilon_m=0$}
\label{Sec.three}

Let us start with the situation \JB{corresponding} to \JB{a} sufficiently long topological
nanowire where the overlap between the Majorana modes is negligible, $\epsilon_{m} \rightarrow 0$. Under such circumstances, $E_{1}^{\pm}=E_{2}^{\pm}$ and $E_{3}^{\pm}=E_{4}^{\pm}$ (nonvanishing overlap $\epsilon_{m}$ lifts this degeneracy). For \JB{a} positive value $\epsilon_{m}>0$, the ground state energy depends on the QD level $\varepsilon_{d}$ and the Coulomb potential $U_{d}$.

\begin{eqnarray}
\min{E_{i}^{-}}=\left\{ \begin{array}{lll}
 E^{-}_{3} \hspace{1mm} &\mbox{\rm for}& \hspace{1mm} \xi_{d} \leq-U_d/2  , \\
 E^{-}_{4} \hspace{1mm} &\mbox{\rm for}& \hspace{1mm} -U_d/2 < \xi_{d}  \leq 0 , \\
 E^{-}_{2} \hspace{1mm} &\mbox{\rm for}& \hspace{1mm} 0 < \xi_{d}\leq U_d/2 , \\
 E^{-}_{1} \hspace{1mm} &\mbox{\rm for}& \hspace{1mm} U_d/2 < \xi_{d} .
\end{array} \right. 
\label{gr}
\end{eqnarray} 
 To characterize the excitation spectrum of our hybrid system\JB{,} it is convenient to introduce the abbreviations.

\begin{eqnarray}
    E_p&=&\sqrt{(\xi_{d}-U_d/2)^2+4t_m^2} \label{eq13} \\
    E_q&=&\sqrt{(\xi_{d}+U_{d}/2)^2+4t_m^2} \label{eq14} 
\end{eqnarray}
and define the  coefficients 
\begin{eqnarray}
    u_{p}^2&=&\frac{1}{2} \left[1+\frac{\xi_{d}-U_d/2}{E_p} \right] =1-v_{p}^2 ,\\
    u_{q}^2&=&\frac{1}{2} \left[1+\frac{\xi_{d}+U_d/2}{E_p} \right] =1-v_{q}^2 .
\end{eqnarray}
For $\epsilon_m=0$ the eigenvectors (\ref{eq4}-\ref{eq7}) simplify to
\begin{eqnarray}
|\Psi^{-}_{1} \rangle &=& u_p|0,0\rangle+v_p|\downarrow,1\rangle ,
\label{eq16} \\ 
|\Psi^{+}_{1} \rangle &=& v_p|0,0\rangle-u_p|\downarrow,1\rangle , 
\label{eq17} \\ 
|\Psi^{-}_{2} \rangle &=&v_p|\downarrow,0\rangle+u_{p} |0,1\rangle ,
\label{eq18} \\ 
|\Psi^{+}_{2} \rangle &=&u_p|\downarrow,0\rangle-v_{p} |0,1\rangle ,
\label{eq19} \\ 
|\Psi^{-}_{3} \rangle &=& v_q |\uparrow \downarrow,0\rangle +u_q|\uparrow,1\rangle ,
\label{eq20} \\ 
|\Psi^{+}_{3} \rangle &=&u_q |\uparrow \downarrow,0\rangle -v_q|\uparrow,1\rangle ,
\label{eq21} \\ 
|\Psi_{4}^{-}\rangle&=& u_q |\uparrow,0\rangle+v_q |\uparrow\downarrow,1\rangle ,
\label{eq22} \\ 
|\Psi_{4}^{+}\rangle&=& v_q |\uparrow,0\rangle-u_q |\uparrow\downarrow,1\rangle .
\label{eq23} 
\end{eqnarray}
Explicit expressions for $\epsilon_{m}\neq 0$ are discussed in Sec.\ \ref{Sec.four}.

From the set of eigenvectors (\ref{eq16}-\ref{eq23}) and eigenenergies (\ref{eq8a}-\ref{eq10}), we can construct \JB{arbitrary} Green's functions, using the spectral Lehmann representation. We assume our setup to be in thermal equilibrium with an external bath, for instance, the substrate on which the topological nanowire is deposited and/or the conducting STM tip.

\subsection{Spectrum of spin-$\downarrow$ electrons}

The Fourier transform of the single-particle propagator of spin-$\downarrow$ electrons is given by
\begin{eqnarray}
    \langle\langle \hat{d}_{\downarrow} ; \hat{d}_{\downarrow}^{\dagger} \rangle \rangle_{\omega}=\frac{1}{Z} \sum_{m,n,s,\bar{s}} |\langle \Psi_{m}^{\bar{s}}|\hat{d}_{\downarrow}| \Psi_{n}^{s} \rangle|^2 \frac{e^{-\beta E_n^{s}}+e^{-\beta E_{m}^{\bar{s}}}}{\omega + E_{n}^{s}-E_{m}^{\bar{s}}} ,
    \label{GF_down}
\end{eqnarray}
where $Z=\sum_{n,s}\exp{\left(-\beta E_{n}^{s}\right)}$ denotes the partition function and
$\beta=(k_{B}T)^{-1}$ is \JB{the} inverse temperature. Indices $m,n =1,2,3,4$ and $s$, $\bar{s} = \pm$ denote particular states introduced in Eqs.\ (\ref{eq16}-\ref{eq23}) as well as their corresponding energies $E_{m/n}^{s/\bar{s}}$ given by Eqs.\ (\ref{eq8a}-\ref{eq10}).
Transitions $\langle \Psi_{m}^{\bar{s}}|\hat{d}_{\downarrow}| \Psi_{n}^{s} \rangle$ are allowed only between the different parity states $\Psi_{1}^{s} \leftrightarrow \Psi_{2}^{\bar{s}}$ and $\Psi_{3}^{s} \leftrightarrow \Psi_{4}^{\bar{s}}$. Contribution to \JB{the} zero-energy mode is given by transitions between degenerate states. Such degeneration occurs between particular states with the same indices ($s=\bar{s}$) as $E_1^{\pm}=E_{2}^{\pm}$ and $E_3^{\pm}=E_{4}^{\pm}$. 
Matrix elements of such transitions are given by $|\langle\Psi_{1}^{+}|\hat{d}_{\downarrow}| \Psi_{2}^{+} \rangle|^2=|\langle\Psi_{1}^{-}|d_{\downarrow}| \Psi_{2}^{-} \rangle|^2=u_{p}^{2}v_{p}^2$ 
and $|\langle\Psi_{3}^{-}| d_{\downarrow}|\Psi_{4}^{+} \rangle|^2=|\langle\Psi_{3}^{+}|\hat{d}_{\downarrow}| \Psi_{4}^{-} \rangle|^2=u_{q}^{2}v_{q}^2$.
Therefore, \JB{the} zero-energy pole contribution to the Green's function can be written as

\begin{equation}
\frac{1}{Z}\sum_{m,n}\sum_{s}|\langle \Psi^{s}_n|\hat{d}_{\downarrow}|\Psi^{s}_m\rangle|^2 \frac{e^{-\beta E_n^{s}}+e^{-\beta E_{m}^{s}}}{\omega + E_{n}^{s}-E_{m}^{s}}  
    =\frac{A_1}{\omega+i0^{+}} 
\end{equation}
with the spectral weight
\begin{eqnarray}
    A_1=\frac{4}{Z}\sum_{s=\pm}  \left[u_p^2v_p^2\left( e^{-\beta E_1^{s}}\right) + u_q^2v_q^2\left( e^{-\beta E_3^{s}} \right) \right] .
\end{eqnarray}

On the other hand, transitions between the states $\Psi^{s}_{1} \leftrightarrow \Psi^{\bar{s}}_{2}$ and $\Psi^{s}_{3} \leftrightarrow \Psi^{\bar{s}}_{4}$ with different sign index $s\neq\bar{s}$ contribute \JB{to} the finite-energy poles at  $\pm E_{p}$ and $\pm E_{q}$, respectively.  
For $\Psi^{s}_{1} \leftrightarrow \Psi^{\bar{s}}_{2}$ matrix elements are given by $|\langle\Psi_{1}^{+}|d_{\downarrow}| \Psi_{2}^{-} \rangle|^2=|\langle\Psi_{2}^{+}|d_{\downarrow}| \Psi_{1}^{-} \rangle|^2=v_{p}^{4}$ 
and $|\langle\Psi_{1}^{-}| d_{\downarrow}|\Psi_{2}^{+} \rangle|^2=|\langle\Psi_{2}^{-}|d_{\downarrow}| \Psi_{1}^{+} \rangle|^2=u_{p}^{4}$. \JB{The} contribution to the Green's function from \JB{the} first two transitions takes \JB{the} form:

\begin{eqnarray}
   &&\frac{1}{Z} \sum_{n,m=1,2}|\langle \Psi^{+}_n|\hat{d}_{\downarrow}|\Psi^{-}_m\rangle|^2 \frac{e^{-\beta E_n^{+}}+e^{-\beta E_{m}^{-}}}{\omega + E_{n}^{+}-E_{m}^{-}} = \nonumber \\ &=& 2\frac{v_{p}^4}{Z}\frac{e^{-\beta E_1^{+}}+e^{-\beta E_{1}^{-}}}{\omega + E_{p}}=\frac{2}{Z}\sum_{s=\pm} v_{p}^4 \frac{e^{-\beta E_{1}^{s}}}{\omega+E_{p}} 
   \label{aux.eq1}
\end{eqnarray}

 Similarly, for the latter two we have
\begin{eqnarray}
    &&\frac{1}{Z} \sum_{n,m=1,2}|\langle \Psi^{-}_n|\hat{d}_{\downarrow}|\Psi^{+}_m\rangle|^2 \frac{e^{-\beta E_n^{-}}+e^{-\beta E_{m}^{+}}}{\omega + E_{n}^{+}-E_{m}^{-}}=  \nonumber \\ &=& \frac{2}{Z}\sum_{s=\pm} u_{p}^4 \frac{e^{-\beta E_{1}^{s}}}{\omega-E_{p}}  
    \label{aux.eq2}
\end{eqnarray}

\JB{The} total contribution to the trivial states from all transitions between $\Psi^{s}_{1} \leftrightarrow \Psi^{\bar{s}}_{2}$ can thus be written as

\begin{eqnarray}
\nonumber
\frac{1}{Z}   &&\sum_{m,n}^{1,2}\sum_{s=\pm} |\langle \Psi^{s}_n|\hat{d}_{\downarrow}|\Psi^{-s}_m\rangle|^2 \frac{e^{-\beta E_n}+e^{-\beta E_{m}}}{\omega + E_{n}-E_{m}} = \\ &=& \frac{A_2}{\omega+E_p+i0^+}+ \frac{A_3}{\omega-E_p+i0^+},
\end{eqnarray}

with amplitudes 
\begin{eqnarray}
    A_2&=&\frac{2}{Z}\sum_{s=\pm}  u_p^4 e^{-\beta E_1^{s}}\\
    A_3&=&\frac{2}{Z}\sum_{s=\pm}  v_p^4 e^{-\beta E_1^{s}}.
\end{eqnarray}
Analogous calculations for $m$ and $n = 3,4$ give
\begin{eqnarray}
\nonumber
    &&\sum_{m,n}^{3,4}\sum_{s=\pm} |\langle \Psi^{s}_n|\hat{d}_{\downarrow}|\Psi^{-s}_m\rangle|^2 \frac{e^{-\beta E_n}+e^{-\beta E_{m}}}{\omega + E_{n}-E_{m}} = \\ &=& \frac{A_4}{\omega+E_q+i0^+}+ \frac{A_5}{\omega-E_q+i0^+}
\end{eqnarray}
with amplitudes
\begin{eqnarray}
    A_4&=&\frac{2}{Z}\sum_{s=\pm}  u_q^4 e^{-\beta E_3^{s}} , \\ 
    A_5&=&\frac{2}{Z}\sum_{s=\pm}  v_q^4 e^{-\beta E_3^{s}} .
\end{eqnarray}

The density of states  $\rho_{\downarrow}(\omega)=-\frac{1}{\pi}\mbox{\rm Im}\langle\langle \hat{d}_{\downarrow} ; \hat{d}_{\downarrow}^{\dagger} \rangle \rangle_{\omega+i0^+}$ of spin-$\downarrow$ electrons  
consists of five branches
\begin{eqnarray}
\nonumber
    \rho_{\downarrow}(\omega)&=&A_1\delta(\omega)+A_2\delta(\omega-E_{p})+A_3\delta(\omega+E_{p})\\ 
    &+&A_{4}\delta(\omega-E_{q})+A_{5}\delta(\omega+E_{q}) ,
    \label{LDOS}
\end{eqnarray}
where $A_{1}$ represents \JB{the} spectral weight of
the Majorana mode transmitted onto the correlated quantum dot, and the amplitudes $A_{2-5}$ refer to the trivial (finite-energy) quasiparticles. 
The coefficients $A_i$ represent the spectral weights of \JB{the} given quasiparticles. In other words, these dimensionless numbers ($A_{i}$) can be regarded as probabilities for the existence of the quasiparticles at the energies $\omega_{i}$.
The total spectral weight satisfies the sum rule
$\sum_{i=1}^{5}A_{i}=1$.

\begin{figure}[b!]
	\includegraphics[width=\wymiarm]{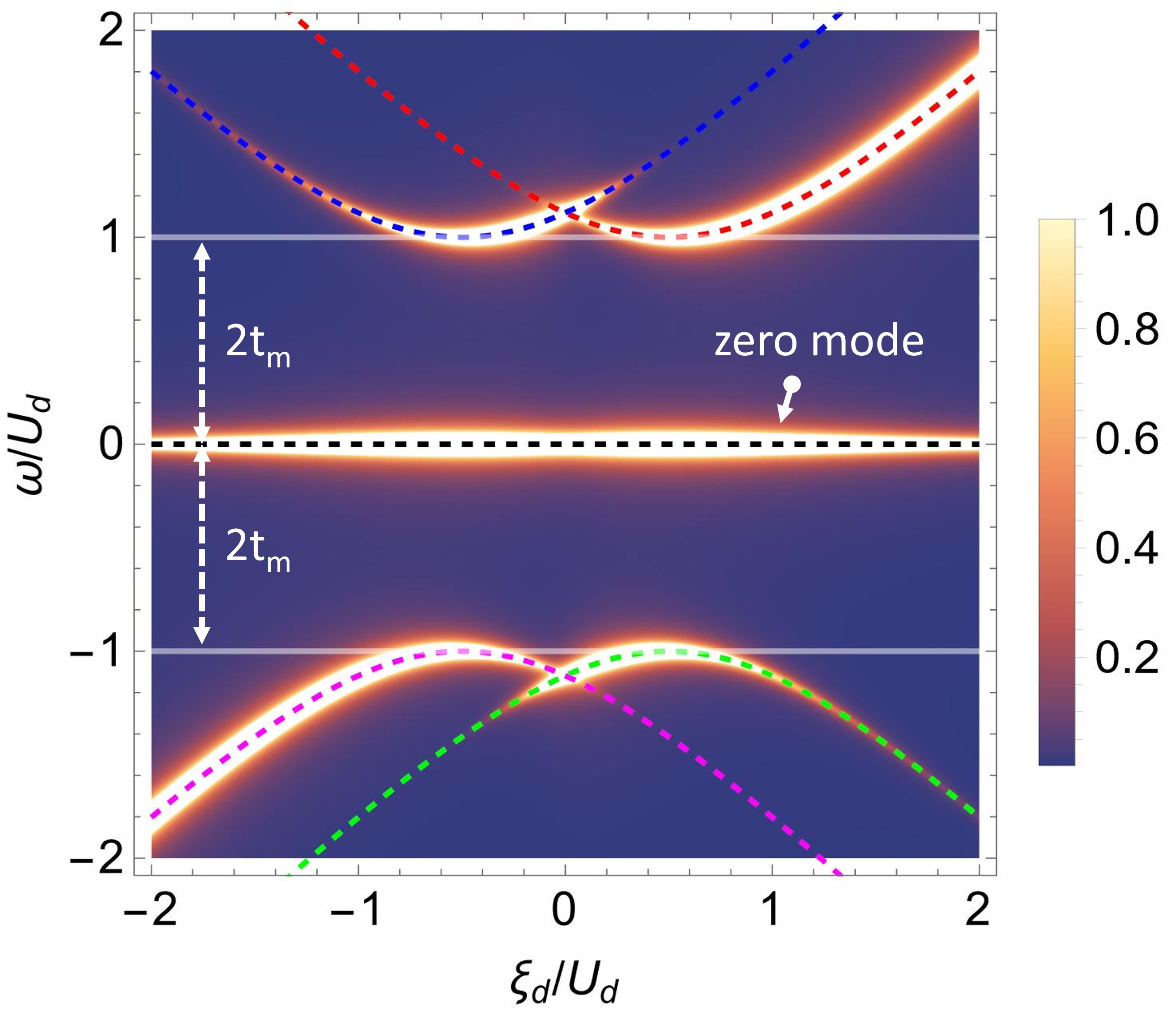}
	\caption{Five quasiparticle branches of the spin-resolved spectrum $\rho_{d\downarrow}(\omega)$ 
\JB{vary} with respect to $\xi_{d}=\varepsilon_d+U_d/2$. Dashed lines show the quasiparticle energies, and their spectral weights, $A_{i}$, are displayed according to \JB{the} r.h.s.\ bar scale. White faded lines indicate \JB{the} topological gap separating ordinary states from the induced zero mode.
} 
	\label{fig5pole}
\end{figure}

Figure \ref{fig5pole} shows the typical spectrum of $\downarrow$-spin electrons. The black dashed line indicates the zero-energy quasiparticle, originating from the Majorana mode leakage. Red/green dashed lines correspond to the quasiparticle energies $\pm E_{p}$ and blue/magenta \JB{indicate} the quasiparticle energies $\pm E_q$, respectively. To understand their physical meaning, let us recall that an isolated QD ($t_m=0$) has two quasiparticle energies: at $\omega=\varepsilon_{d}$ (i.e. $\xi_d=-\frac{U_d}{2}$) with spectral weight $1-n_{d\sigma}$ and another Coulomb satellite at $\omega=\varepsilon_{d}+U_{d}$ (i.e. $\xi_d=\frac{U_d}{2}$) with spectral weight $n_{d\sigma}$. For $t_{m}\neq 0$, these quasiparticle branches evolve into the trivial modes $\pm E_{p}$ and $\pm E_{q}$ of our setup, which are \JB{gapped} due to the intersite pairing (for details see Sec.\ \ref{section_pairing}). 
In Figure \ref{fig5pole}, we clearly notice {\it avoided-crossing} behavior of the trivial (finite-energy) quasiparticle branches, repelled at some distance from the topological (zero-energy) mode, which is due to the protection of the Majorana state. At the critical points, $\xi_d = \pm \frac{U_d}{2}$, the trivial states are separated from the zero-energy mode by a gap of $2t_m$.
Furthermore, the spectral weight $A_1$ of the zero-energy mode, $\omega=0$, is enhanced around 
$\omega=\xi_d-\frac{U_d}{2}$ and 
$\omega=\xi_{d}+\frac{U_d}{2}$.

\begin{figure}
	\includegraphics[width=\wymiarm]{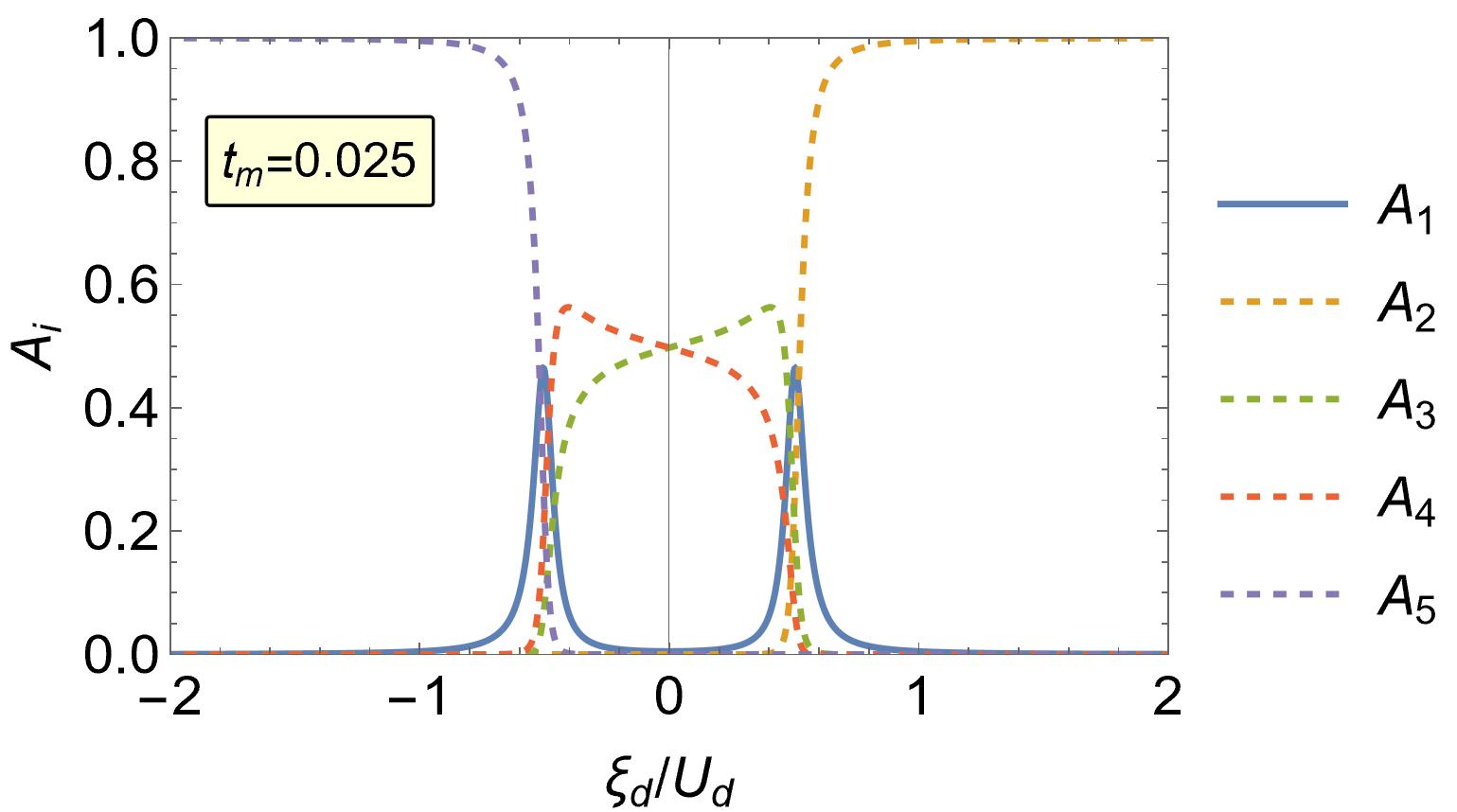}
	\includegraphics[width=\wymiarm]{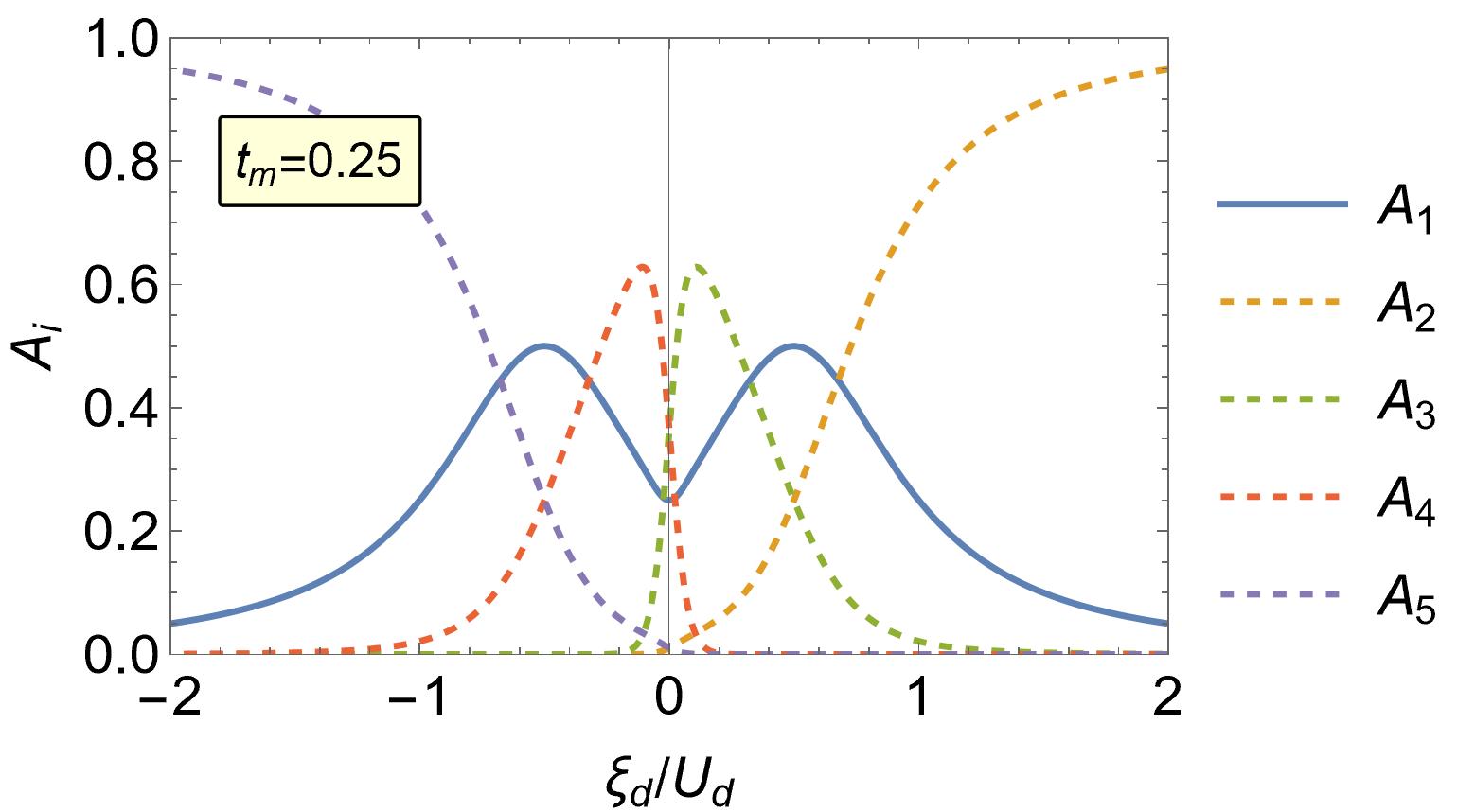}
	\includegraphics[width=\wymiarm]{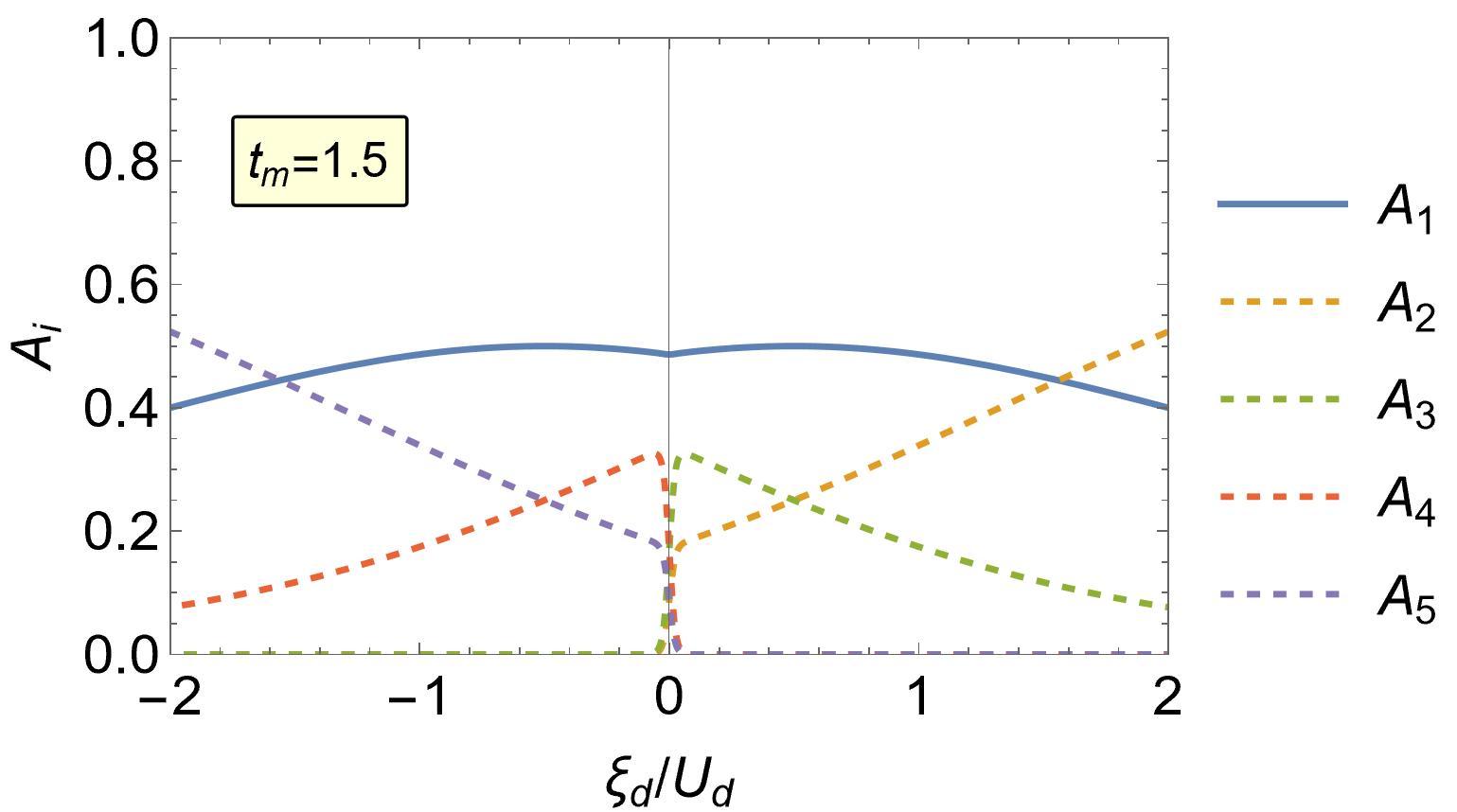}
 \caption{Variation of the spectral weights $A_{1-5}$ against the QD energy level obtained for \JB{the} weak coupling $t_{m}/U_{d}=0.025$ (top panel), intermediate hybridization $t_{m}/U_{d}=0.25$ (middle panel), and in the strong coupling limit $t_{m}/U_{d}=1.5$ (bottom panel).
} 
	\label{fig.tm1}
\end{figure}

\begin{figure}
	\includegraphics[width=\wymiarm]{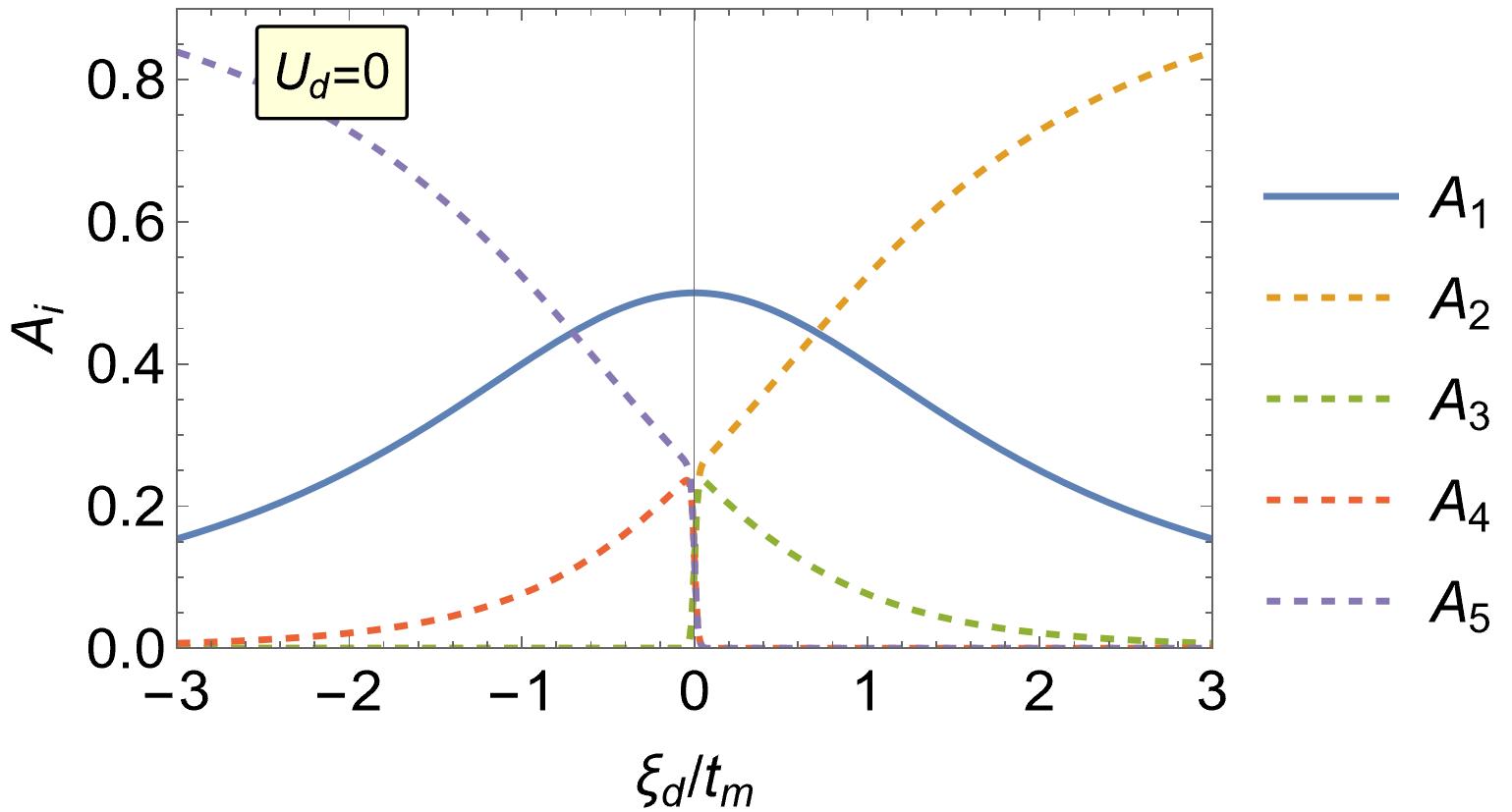}
	\includegraphics[width=\wymiarm]{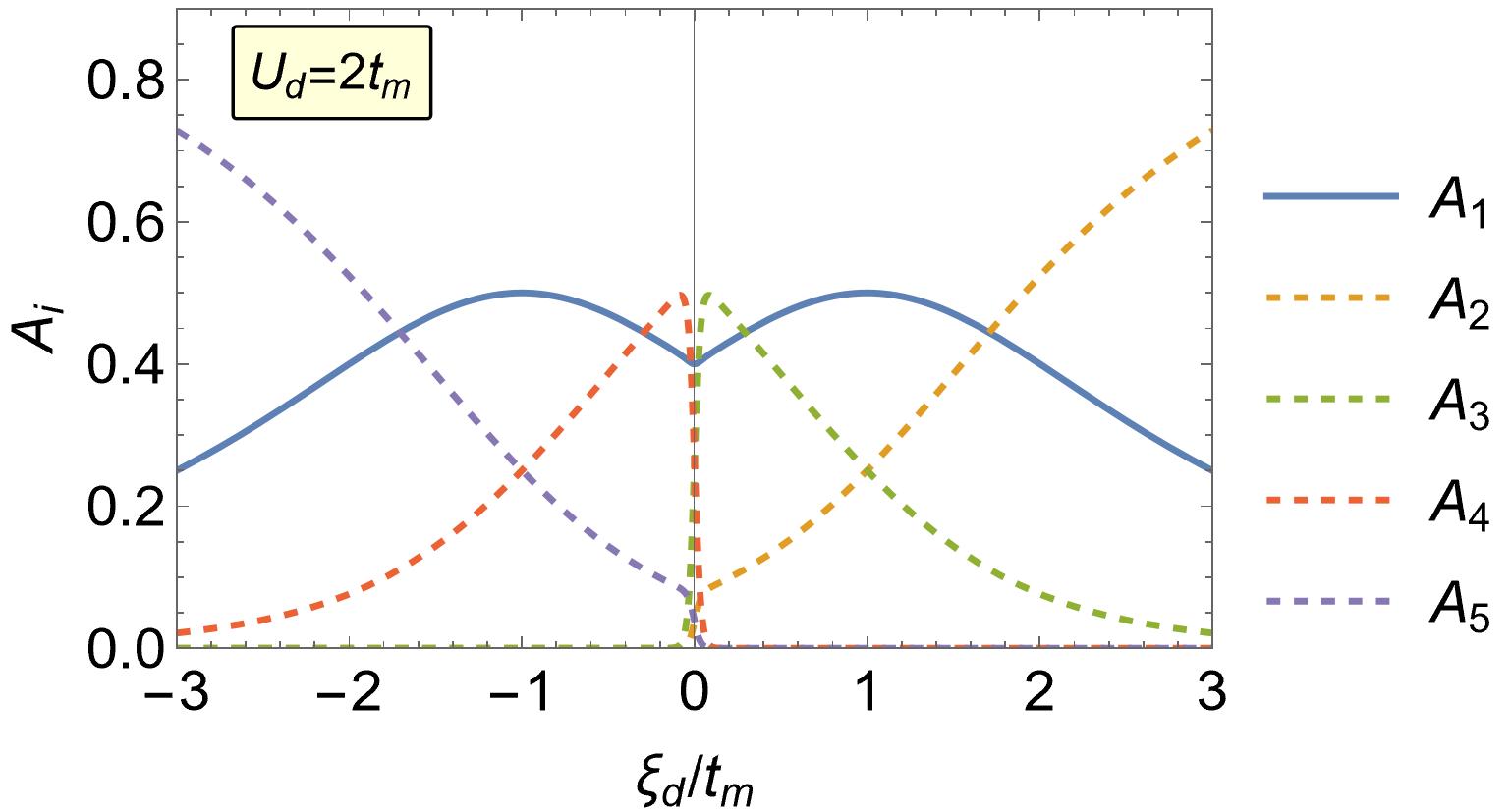}
	\includegraphics[width=\wymiarm]{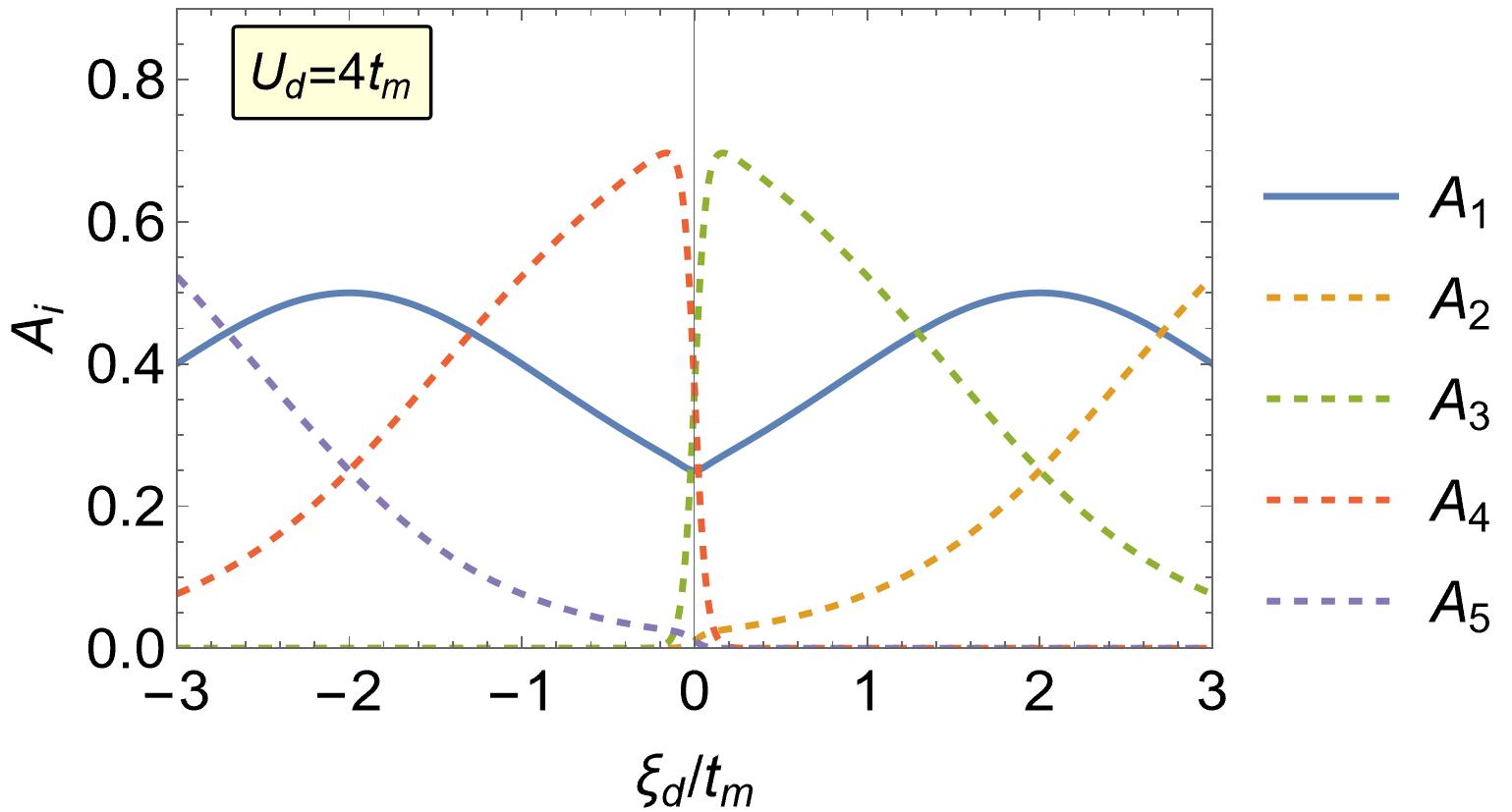}
 \caption{Variation of the spectral weights $A_{i}$ with respect to the quantum dot energy level $\xi_{d}=\varepsilon_{d}+U_{d}/2$ obtained for several values of the Coulomb potential $U_d$, as \JB{indicated}.
} 
	\label{fig.tm2}
\end{figure}

To specify the optimal spectral weight of the Majorana mode, we present in Figs.\ \ref{fig.tm1} and \ref{fig.tm2} the variation of all amplitudes $A_{i}$ against $\xi_{d}$. These plots demonstrate \JB{that}, for the weak coupling $t_m$, practically only two trivial quasiparticles coexist with the zero-energy mode. In other words, the spectrum of $\downarrow$-spin electrons exhibits three dominant (out of five) quasiparticle branches. As regards the zero-energy mode, its optimal spectral weight coincides with $\xi_d=-\frac{U_d}{2}$ and $\xi_{d}=\frac{U_d}{2}$.


Figure \ref{fig.tm1} shows \JB{the} influence of the coupling $t_m$ on \JB{the} energy region in which the Majorana spectral weight is noticeable. For infinitesimal coupling 
$t_m$, the Majorana mode exists only very close to the quasiparticle energies $\xi_d \pm \frac{U_d}{2}$. Upon increasing $t_{m}$, the Majorana mode extends onto a much broader region around those energies. In the case of very strong dot-Majorana coupling ($t_m>U_d$), the quantum dot can be considered as an additional atom embedded in the topological chain. For such a "molecular" case, leakage of MZM is efficient over a wide range of $\xi_{d}$ (c.f. bottom panel of Fig.\ \ref{fig.tm1}).
Let us remark that the optimal value, $\max{\left\{A_{1}\right\}}=0.5$, coincides with the minima of $|E_{q,p}|$.

Similar behavior is observed when inspecting the influence of the Coulomb potential $U_d$, Fig.\ \ref{fig.tm2}. In particular, at half-filling 
($\xi_d=0$), the spectral weight of the Majorana mode approaches its optimal value only for vanishing Coulomb repulsion $U_d \rightarrow 0$. For stronger Coulomb potential, the optimal spectral weight of the Majorana mode shifts from half-filling (as can be observed in the density of states, Fig.\ \ref{fig5pole}). We have checked that for $U_d=4t_m$, the quasiparticle spectral weights at half-filling acquire the following values: $A_1= 0.25$, $A_2=A_5\simeq 0.01$, $A_3=A_4\simeq 0.0365$.
At half-filling, the effectiveness of MZM leakage diminishes with increasing correlation strength $U_{d}$. The maximal value of the spectral weight of $A_1$ (reaching 0.5) is preserved upon strong correlations when  
$\xi_d=\pm\frac{U_d}{2}$.

\subsection{Spectrum of spin-$\uparrow$ electrons}

The excitation spectrum of $\uparrow$-electrons reveals qualitatively different \JB{behavior}, even though the interaction term, $U_d \hat{n}_{\uparrow}\hat{n}_{\downarrow}$, mixes both spin sectors.
The single-particle Green's function $\langle \langle \hat{d}_{\uparrow} ; \hat{d}_{\uparrow}^{\dagger}\rangle \rangle_{\omega}$ can be expressed in \JB{a} form analogous to Eq.\ (\ref{GF_down}). The only nonvanishing matrix elements are contributed by $\langle \Psi_{2}^{s} | \hat{d}_{\uparrow}| \Psi_{3}^{s'} \rangle$ and $\langle \Psi_{1}^{s} | \hat{d}_{\uparrow}| \Psi_{4}^{s'} \rangle$. 
We note that for $\epsilon_m=0$ the pairs of quasiparticles appearing in these elements are not degenerate. For this reason\JB{,} we observe four branches of the trivial quasiparticles (instead of two typical for the isolated QD). The spectrum of $\uparrow$ electrons does not show \JB{the} presence of the Majorana mode, which should be pinned to zero energy.
Degenerate pairs of the eigenstates $\Psi_{1}^{\pm}$, $\Psi_{2}^{\pm}$\JB{,} and $\Psi_{3}^{\pm}$, $\Psi_{4}^{\pm}$ (in the case $\epsilon_m=0$) imply that
components of the Green function obtained from the matrix elements $\langle \Psi_{2}^{s} | \hat{d}_{\uparrow}| \Psi_{3}^{s'} \rangle$ are identical \JB{to} those originating from $\langle \Psi_{1}^{s} | \hat{d}_{\uparrow}| \Psi_{4}^{s'} \rangle$. Accordingly, we obtain the following four-pole structure of the Green's function:

\begin{eqnarray}
\nonumber
    \langle \langle \hat{d}_{\uparrow} ; \hat{d}_{\uparrow}^{\dagger}\rangle \rangle_{\omega} &=&  \frac{B_{1}}{\omega-\xi_{d}+\frac{1}{2}(E_{p}+E_{q})} \nonumber \\
   &+& \frac{B_{2}}{\omega-\xi_{d}-\frac{1}{2}(E_{p}+E_{q})} \nonumber \\
   &+& \frac{B_{3}}{\omega-\xi_{d}+\frac{1}{2}(E_{p}-E_{q})} \nonumber \\
   &+& \frac{B_{4}}{\omega-\xi_{d}-\frac{1}{2}(E_{p}-E_{q})}, 
\end{eqnarray}
with the amplitudes
\begin{eqnarray}
    B_{1}&=&\frac{2}{Z}(v_{p}u_{q}-u_{p}v_{q})^2 (e^{-\beta E_{1}^{+}}+e^{-\beta E_{3}^{-}}) ,\\
    B_{2}&=&\frac{2}{Z}(u_{p}v_{q}-v_{p}u_{q})^2 (e^{-\beta E_{1}^{-}}+e^{-\beta E_{3}^{+}}) ,\\
    B_{3}&=&\frac{2}{Z}(u_{p}u_{q}+v_{p}v_{q})^2 (e^{-\beta E_{1}^{+}}+e^{-\beta E_{3}^{+}}) , \\
    B_{4}&=&\frac{2}{Z}(u_{p}u_{q}+v_{p}v_{q})^2 (e^{-\beta E_{1}^{-}}+e^{-\beta E_{3}^{-}}) .
\end{eqnarray}
In the energy region $\varepsilon_{d} \in (-U_d,0)$, two amplitudes $B_3$ and $B_4$ are negligibly small, so the dominant contribution comes from $B_{1}$ and $B_{2}$. Outside of this regime, the prevailing contributions are from $B_{3}$ and $B_4$.

\begin{figure}
	\includegraphics[width=\wymiarm]{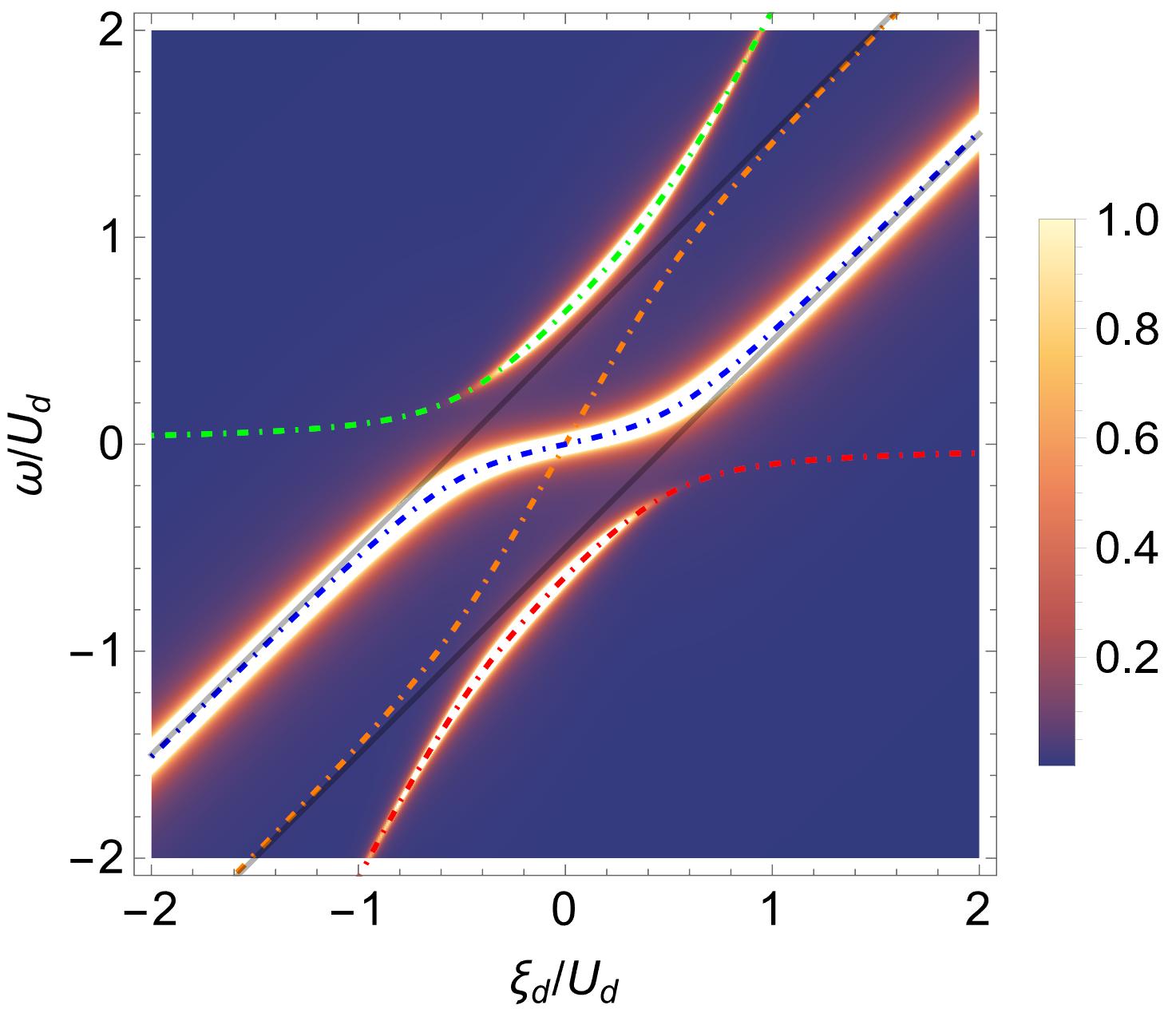}
	\caption{Variation of the quasiparticle spectrum 
$\rho_{d\uparrow}(\omega)$ with respect to $\xi_{d}=\varepsilon_d+U_d/2$ obtained for $t_m=0.2U_d$. The dashed lines mark the positions of four poles, and their spectral weights are displayed by \JB{color-width}, whose scale is indicated by the r.h.s.\ bar. Black faded lines mark the position of two poles $\omega=\xi_d\pm\frac{U_d}{2}$, which remain in the case $t_m=0$.
 } 
	\label{dos_up}
\end{figure}

Figure \ref{dos_up} displays the spectrum of $\uparrow$-spin electrons obtained for the same set of parameters as in Fig.\ \ref{fig5pole}. We clearly notice \JB{the} absence of the Majorana mode. 
Although $\uparrow$ electrons are not directly coupled to the MZM in the considered model, MZM leakage to $\downarrow$ electrons affects the opposite spin spectrum through electron correlations $(U_d)$. Comparing the obtained results to the case where the MZM is completely absent ($t_m = 0$), we observe that instead of two ordinary states located at $\omega = \xi_d \pm \frac{U_d}{2}$, we obtain four branches.
The most pronounced branch, represented by the blue dashed line in Fig.\ \ref{dos_up} (corresponding to transitions between the states $\Psi_{1}^{-} \leftrightarrow \Psi_{3}^{-}$ and $\Psi_{2}^{-} \leftrightarrow \Psi_{4}^{-}$), reproduces the state located at $\omega = \xi_d - \frac{U_d}{2}$ for fillings \JB{way above} 0.5 (i.e., $\xi_d \ll 0$) and the state at $\omega = \xi_d + \frac{U_d}{2}$ in the opposite case. Near half-filling, this state crosses the zero-energy level. The branch represented by the orange line behaves in the opposite manner, crossing zero energy under the same conditions, but with an inverse dependence on the filling.
Two quasiparticle branches crossing at zero energy for half-filling, $\xi_d=0$, \JB{have} nothing to do with \JB{the} topological mode. \JB{The} influence of \JB{the} topological superconductor is merely responsible for doubling the initial branches $\xi_{d}\pm U_{d}/2$ and for interconnecting two of them (internal ones).

\subsection{Magnetization}

\begin{figure}[h]
	\includegraphics[width=\wymiarm]{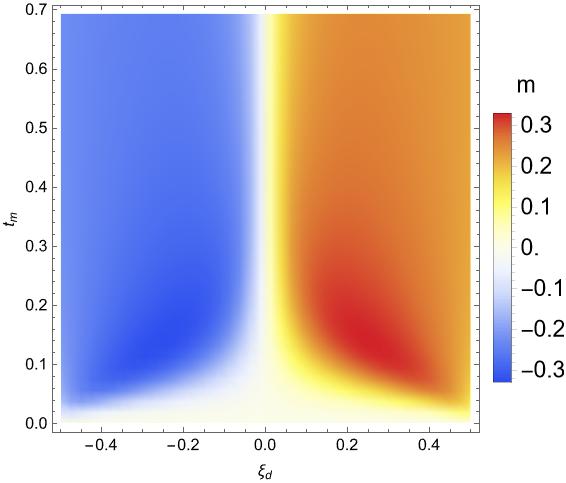}
	\caption{Magnetization of QD as function of $\xi_{d}=\varepsilon_{d}+\frac{U_d}{2}$ and the hybridization stregth $t_{m}$.} 
	\label{magnetization}
\end{figure}

Qualitative differences of the opposite spin spectra are indirectly manifested by the on-dot magnetization 
\begin{eqnarray}
    m=\frac{1}{2}(\langle n_{\downarrow} \rangle - \langle n_{\uparrow} \rangle) .
    \label{magnetization_eq}
\end{eqnarray}
emerging outside the half-filling (see Fig.\ \ref{magnetization}). To explain the sign change of QD magnetization, let us inspect Eqs.\ (\ref{eq4}-\ref{eq7}), noting that $\Psi_{1}$ and $\Psi_2$ represent superpositions of the empty and singly occupied spin-$\downarrow$ configurations. Therefore, if for \JB{particular} model parameters state $\Psi_{1}$ or $\Psi_{2}$ is the ground state, the dot magnetization would be oriented along \JB{the} $\downarrow$-direction. In contrast, the states $\Psi_{3}$ and $\Psi_{4}$ are superpositions of the nonmagnetic $|\uparrow \downarrow\rangle$ state and the singly occupied spin-$\uparrow$ configuration. The ground state represented by $\Psi_{3}$ and $\Psi_{4}$ would then have magnetization along \JB{the} $\uparrow$-direction. Fig.\ 2 shows that for $\xi_{d}<0$, the ground state of QD is represented by $\Psi_{3}$ or $\Psi_{4}$. This fact explains the sign change of the magnetization at half-filling, $\xi_d=0$, in agreement with previous studies reported in Refs.\cite{Lopez-2013,Lutchyn-2014,Huguet2023}.

    \begin{figure}[b!]
	\includegraphics[width=\wymiarm]{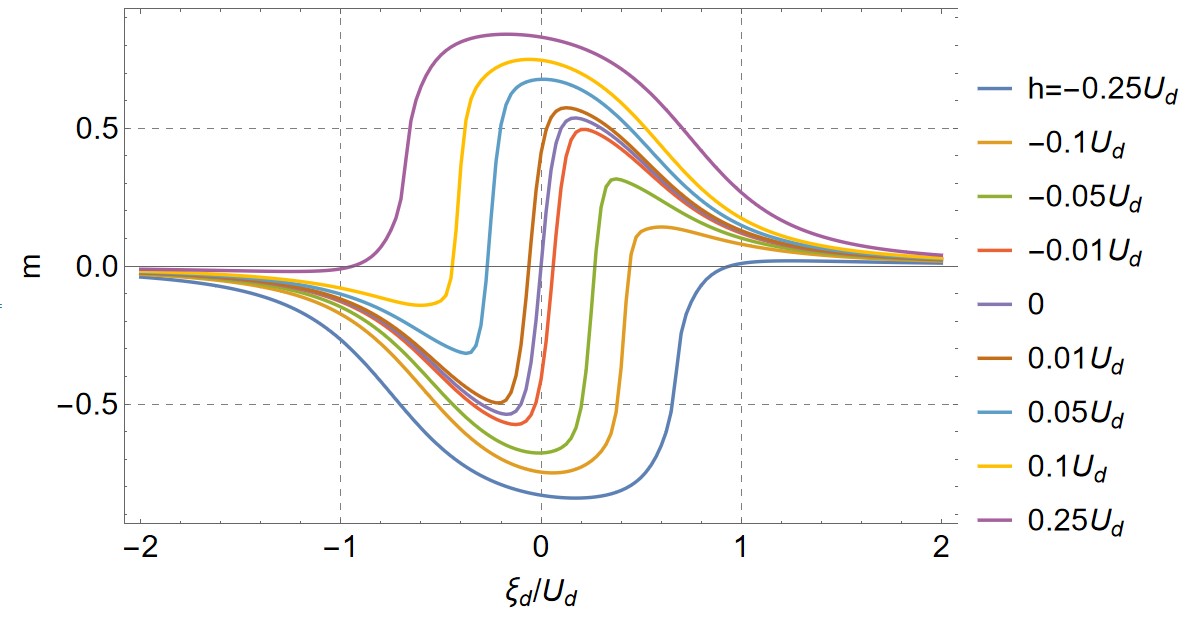}
	\caption{Magnetization of QD as function of $\xi_{d}=\varepsilon_{d}+\frac{U_d}{2}$ in presence of Zeeman field $h$ obtained for $t_m=0.25U_d$.} 
	\label{magnetization_zem}
\end{figure}

As shown in the Appendix (Fig.\ \ref{fig.zemane}), the Zeeman field modifies the eigenenergies in such a way that, when aligned with the spin-down state, the crossing point between energies $E_{12}$ and $E_{34}$ shifts toward lower values of $\xi_d$ (energies become equal at smaller $\xi_d$ compared to the case without the field). This shift affects the polarization transition point of the magnetization. Specifically, for weak Zeeman fields (particularly when aligned with the spin-down orientation), the system favors spin-down polarization over a broader range of energies. Consequently, the magnetization turning point also occurs at lower energy levels $\xi_d$, as illustrated in Figure \ref{magnetization_zem}.
As the Zeeman field strength increases and surpasses a critical value (for $t_m = 0.25U_d$, this occurs at approximately $h > 0.25U_d$), the quantum dot's magnetization becomes polarized in a single direction over a wide range of energy levels $\epsilon_{d}$. The results indicate that, in the presence of a strong Zeeman field, the only region where the quantum dot exhibits significant magnetization is when the magnetization is fully polarized in one direction. The transition point between opposite polarizations occurs at an energy where, beyond this point, the magnetization becomes very small. Therefore, in the region where the magnetization is substantial, it is aligned in a single direction. At this point, the system enters a regime where the external magnetic field dominates \JB{over the} Majorana leakage influence, enforcing a rigid spin alignment.

\subsection{Signatures of intersite pairing}
\label{section_pairing}

The usual method for probing the QD quasiparticle spectrum relies on charge tunneling induced upon applying voltage between our setup \JB{and an} external conducting tip. This sort of measurement has been reported by Deng {\it et al.} \cite{Deng-2018}, revealing enhancement of the zero-bias conductance. 

\begin{figure}
	\includegraphics[width=\wymiarm]{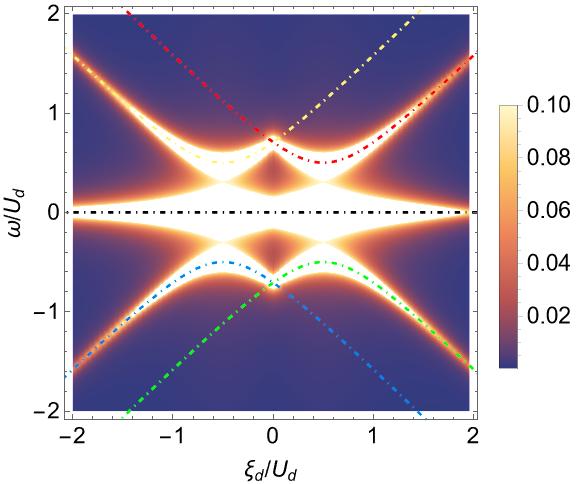}
	\includegraphics[width=\wymiarm]{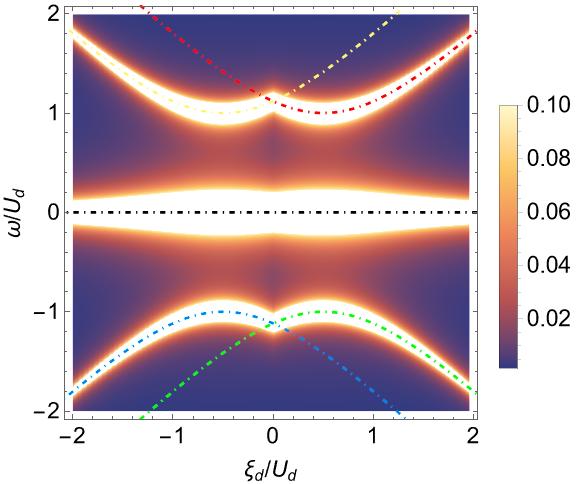}
	\includegraphics[width=\wymiarm]{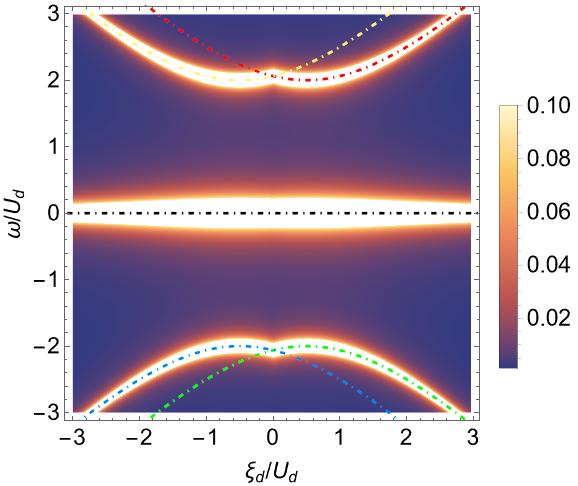}
 \caption{Transmittance of the selective equal spin Andreev reflection (SESAR) obtained for $t_m/U_{d}=0.25$ (top panel), $0.5$ (middle panel), and $1$ (bottom panel), assuming temperature $k_bT=0.01U_d$ and coupling $\Gamma_{N}=0.01U_d$.} 
	\label{fig.SESAR}
\end{figure}

Another method, proposed in Ref.\ \cite{SESAR-2014}, is based on equal spin Andreev scattering to detect efficiency of converting a given spin electron into a hole of the same polarization. The energy-dependent transmittance via such transport channel is given by $T_{\sigma}(\omega)= \Gamma_{N}^{2} \left( \left| \langle\langle \hat{d}_{\sigma}; \hat{f} \rangle\rangle_{\omega} \right|^{2} + \left| \langle\langle \hat{f} ; \hat{d}_{\sigma} \rangle\rangle_{\omega} \right|^{2} \right)$, where $\Gamma_{N}$ denotes the coupling of QD to the polarized conducting electrode. In the simplest approach, the influence of such an external reservoir would be responsible for a level-broadening. Under such conditions, we can express

\begin{eqnarray}
    \langle\langle \hat{d}_{\downarrow}; \hat{f} \rangle\rangle_{\omega} &=& \frac{1}{Z} \sum_{m,n,s,s'} \langle \Psi_{n}^{s} | \hat{d}_{\downarrow}| \Psi_{m}^{s'} \rangle \langle \Psi_{m}^{s'} | \hat{f}| \Psi_{n}^{s} \rangle
    \nonumber \\
& \times &    \frac{e^{-\beta E_{m}^{s}}+e^{-\beta E_{n}^{s}}}{(\omega +i\Gamma_{N}) + E_{n}^{s}-E_{m}^{s}}
\label{GF_intersite}
\end{eqnarray}
which accounts for the inter-site pairing of $\downarrow$-spin electrons. This Green's function (\ref{GF_intersite}) has the same poles as the single-particle propagator (\ref{GF_down}), but with different spectral weights. The matrix elements involving the states $\Psi_1$ and $\Psi_2$ that contribute to the zero-energy poles are given by
\begin{eqnarray}
  \langle \Psi_{1}^{-} | \hat{d}_{\downarrow}| \Psi_{2}^{-} \rangle \langle \Psi_{2}^{-} | \hat{f}| \Psi_{1}^{-} \rangle =u_p v_p^3 \\
     \langle \Psi_{1}^{+} | \hat{d}_{\downarrow}| \Psi_{2}^{+} \rangle \langle \Psi_{2}^{+} | \hat{f}| \Psi_{1}^{+} \rangle =-u_p^3 v_p \\
     \langle \Psi_{2}^{-} | \hat{d}_{\downarrow}| \Psi_{1}^{-} \rangle \langle \Psi_{1}^{-} | \hat{f}| \Psi_{2}^{-} \rangle =u_p^3 v_p \\
     \langle \Psi_{2}^{+} | \hat{d}_{\downarrow}| \Psi_{1}^{+} \rangle \langle \Psi_{1}^{+} | \hat{f}| \Psi_{2}^{+} \rangle =-u_p v_p^3 
 \end{eqnarray}
In a similar manner, the matrix elements involving $\Psi_3$ and $\Psi_4$ follow the same structure, with $u_p$ and $v_p$ replaced by $u_q$ and $v_q$. Substituting the explicit form of $v_{p(q)}$ and $u_{p(q)}$, the sum of given matrix elements \JB{simplifies} to $u_{p(q)}^3v_{p(q)}+u_{p(q)}v_{p(q)}^3=\frac{t_m}{E_{p(q)}}$.
This leads to the following expression describing the amplitude of the zero-energy pole of the discussed Green's function
\begin{eqnarray}
   \frac{1}{Z} \left[ \frac{2t_{m}}{E_{p}}(e^{-\beta E_{1}^{-}}-e^{-\beta E_{1}^{+}})+ \frac{2t_{m}}{E_{q}}(e^{-\beta E_{3}^{-}}-e^{-\beta E_{3}^{+}}) \right] .
\end{eqnarray}
The finite-energy poles of $ \langle\langle \hat{d}_{\sigma}; \hat{f} \rangle\rangle_{\omega} $ are given by
%
%
\begin{eqnarray}
    \pm\frac{1}{Z} \frac{2t_m}{E_{p}}(e^{-\beta E_{1}^-}+e^{-\beta E_{1}^+})\frac{1}{\omega +i\Gamma_{N} \pm E_p},
\end{eqnarray}
\begin{eqnarray}
    \pm\frac{1}{Z} \frac{2t_m}{E_{q}}(e^{-\beta E_{3}^-}+e^{-\beta E_{3}^+})\frac{1}{\omega +i\Gamma_{N} \pm E_q} .
\end{eqnarray}
%

\JB{A} typical plot of the spin-$\downarrow$ selective Andreev transmittance is presented in Fig.\ \ref{fig.SESAR} for several ratios $t_{m}/U_{d}$. These plots provide clear indication that zero-bias conductance of SESAR is able to probe the spectral weight of the Majorana mode \JB{as it varies} against $\xi_{d}$. Again, we notice that the Coulomb repulsion shifts the optimal weight of such

\section{Results for $\epsilon_m \neq 0$}
\label{Sec.four}

\JB{The} local solution allows us to identify the origin of the quasiparticle spectrum of QD, assigning its specific features to the topological or trivial components. Such {\it identification} becomes a bit more complicated when $\epsilon_m \neq 0$.
In such \JB{a} situation, the eigenstates of our setup $\Psi^{s}_{i}$ are nondegenerate, with the corresponding energies (\ref{eq8a}-\ref{eq10}).
In analogy to the quasiparticle energies (\ref{eq13},\ref{eq14}), it is convenient to define

\begin{eqnarray}
    E^{\pm}_p&=&\sqrt{(\varepsilon_d \pm \epsilon_m)^2+4t_m^2} \label{eq44} \\
    E^{\pm}_q&=&\sqrt{(\varepsilon_d \pm \epsilon_m+U_d)^2+4t_m^2} . \label{eq45}
\end{eqnarray}
which \JB{helps} us to express the coefficients $u^{s}_{i}$ appearing in the eigenstates $|\Psi_{i}^{s}\rangle$ by

\begin{eqnarray}
(u_1^{\pm})^2 &=& \frac{1}{2}\left(1 \pm \frac{\varepsilon_d + \epsilon_m}{E_p^+}\right) , \\
(u_2^{\pm})^2 &=& \frac{1}{2}\left(1 \pm \frac{\varepsilon_d - \epsilon_m}{E_p^-}\right) ,\\
(u_3^{\pm})^2 &=& \frac{1}{2}\left(1 \pm \frac{\epsilon_m - \varepsilon_d + U_d}{E_q^-}\right) , \\
(u_4^{\pm})^2 &=& \frac{1}{2}\left(1 \pm \frac{\varepsilon_d + \epsilon_m + U_d}{E_q^+}\right) ,
\end{eqnarray}
and $(v^{s}_{i})^2=1-(u^{s}_{i})^2$.\\

After algebraic calculations we obtain the following spectral function 
for arbitrary $\epsilon_m$ 
\begin{eqnarray}
    \rho_{\downarrow}(\omega) 
    &=& A_{12}^{-/-}\delta\left[\omega + \frac{1}{2}(E^{-}_p - E^{+}_p)\right] \nonumber \\
    &+& A_{12}^{-/+}\delta\left[\omega - \frac{1}{2}(E^{-}_p + E^{+}_p)\right] \nonumber \\
&+& A_{12}^{+/-}\delta\left[\omega + \frac{1}{2}(E^{-}_p + E^{+}_p)\right]  \nonumber \\
&+& A_{12}^{+/+}\delta\left[\omega - \frac{1}{2}(E^{-}_p - E^{+}_p)\right] \nonumber \\
&+& A_{34}^{-/-} \delta\left[\omega - \frac{1}{2}(E^{-}_q - E^{+}_q)\right] \nonumber \\
&+& A_{34}^{-/+} \delta\left[\omega - \frac{1}{2}(E^{-}_q + E^{+}_q)\right] \nonumber \\
&+& A_{34}^{+/-} \delta\left[\omega + \frac{1}{2}(E^{-}_q + E^{+}_q)\right] \nonumber \\
&+& A_{34}^{+/+} \delta\left[\omega + \frac{1}{2}(E^{-}_q - E^{+}_q)\right]
\end{eqnarray}
with the amplitudes $A_{12}^{s/s'}$ related to transitions $\Psi^{s}_{1} \leftrightarrow \Psi^{s'}_{2}$ \JB{given by}
\begin{eqnarray}
    A^{-/-}_{12}&=&\frac{1}{Z}(u_{1}^{-}u_{2}^{-})^2(e^{-\beta E^{-}_{1}}+e^{-\beta E^{-}_{2}})  \nonumber \\ &+&\frac{1}{Z}(v_{1}^{+}v_{2}^{+})^2(e^{-\beta E^{+}_{1}}+e^{-\beta E^{+}_{2}}) \nonumber ,\\
    A^{-/+}_{12}&=&\frac{1}{Z}(u_{1}^{-}u_{2}^{+})^2(e^{-\beta E^{-}_{1}}+e^{-\beta E^{+}_{2}}) \nonumber \\ &+&\frac{1}{Z}(v_{1}^{-}v_{2}^{+})^2(e^{-\beta E^{-}_{1}}+e^{-\beta E^{+}_{2}}) ,\nonumber \\
    A^{+/-}_{12}&=&\frac{1}{Z}(u_{1}^{+}u_{2}^{-})^2(e^{-\beta E^{+}_{1}}+e^{-\beta E^{-}_{2}}) \nonumber \\ &+&\frac{1}{Z}(v_{1}^{+}v_{2}^{-})^2(e^{-\beta E^{+}_{1}}+e^{-\beta E^{-}_{2}}) ,\nonumber \\
    A^{+/+}_{12}&=&\frac{1}{Z}(u_{1}^{+}u_{2}^{+})^2(e^{-\beta E^{+}_{1}}+e^{-\beta E^{+}_{2}}) \nonumber \\ &+&\frac{1}{Z}(v_{1}^{-}v_{2}^{-})^2(e^{-\beta E^{-}_{1}}+e^{-\beta E^{-}_{2}}) 
\end{eqnarray}
and amplitudes $A_{34}^{s/s'}$ of transitions $\Psi^{s}_{3} \leftrightarrow \Psi^{s'}_{4}$ \JB{given by}
\begin{eqnarray}
    A^{-/-}_{34} &=& \frac{1}{Z}(u_{3}^{-}u_{4}^{-})^2(e^{-\beta E^{-}_{3}} + e^{-\beta E^{-}_{4}}) \nonumber \\
    &+& \frac{1}{Z}(v_{3}^{+}v_{4}^{+})^2(e^{-\beta E^{+}_{3}} + e^{-\beta E^{+}_{4}}) ,\nonumber \\
    A^{-/+}_{34} &=& \frac{1}{Z}(u_{3}^{-}u_{4}^{+})^2(e^{-\beta E^{-}_{3}} + e^{-\beta E^{+}_{4}}) \nonumber \\
    &+& \frac{1}{Z}(v_{3}^{-}v_{4}^{+})^2(e^{-\beta E^{-}_{3}} + e^{-\beta E^{+}_{4}}) , \nonumber \\
    A^{+/-}_{34} &=& \frac{1}{Z}(u_{3}^{+}u_{4}^{-})^2(e^{-\beta E^{+}_{3}} + e^{-\beta E^{-}_{4}}) \nonumber \\
    &+& \frac{1}{Z}(v_{3}^{+}v_{4}^{-})^2(e^{-\beta E^{+}_{3}} + e^{-\beta E^{-}_{4}}) , \nonumber \\
    A^{+/+}_{34} &=& \frac{1}{Z}(u_{3}^{+}u_{4}^{+})^2(e^{-\beta E^{+}_{3}} + e^{-\beta E^{+}_{4}}) \nonumber \\
    &+& \frac{1}{Z}(v_{3}^{-}v_{4}^{-})^2(e^{-\beta E^{-}_{3}} + e^{-\beta E^{-}_{4}}) .
\end{eqnarray}

\begin{figure}[h]
	\includegraphics[width=\wymiarm]{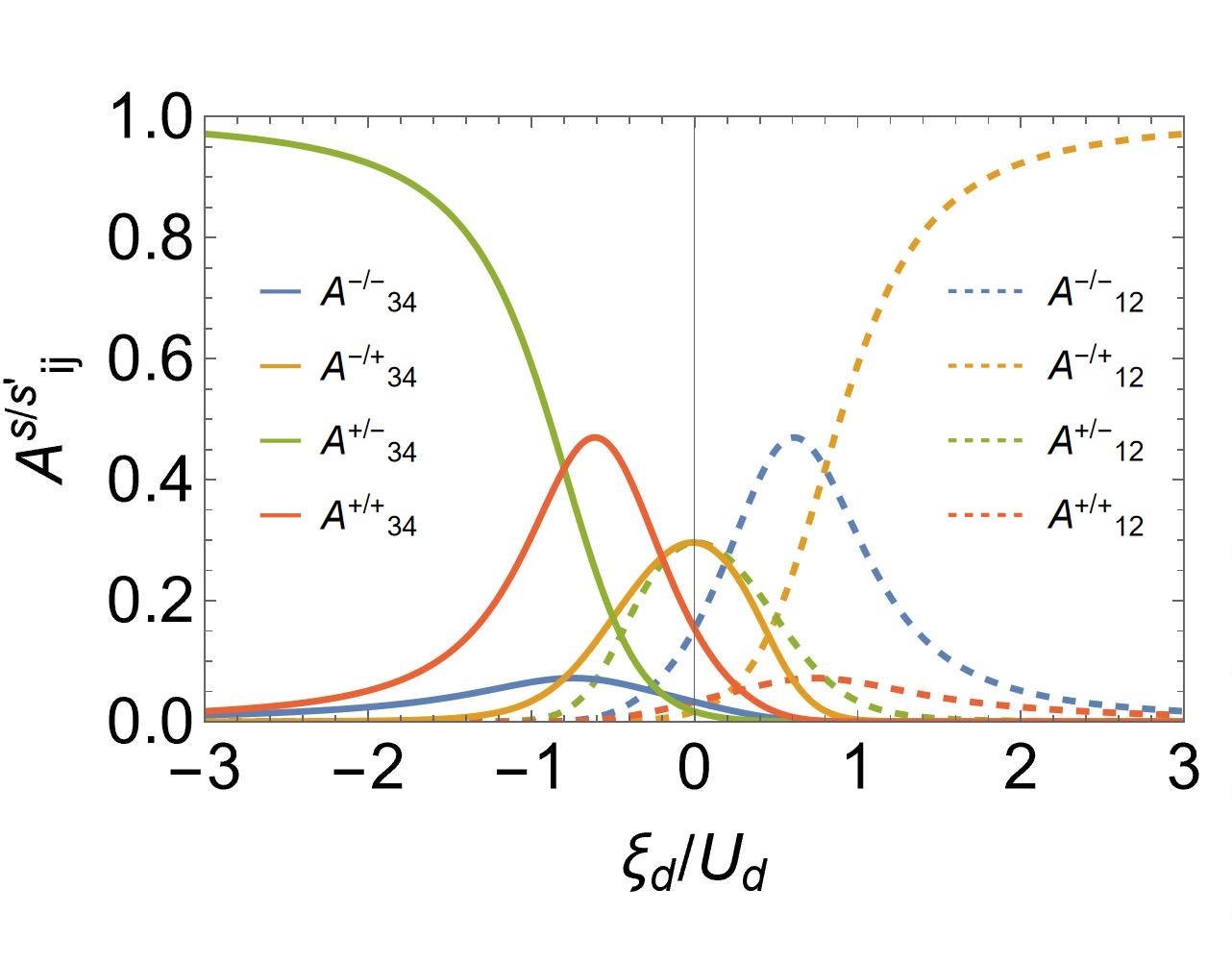}
	\caption{Variation of the transition probabilities $A^{s/s'}_{i j}$ between $\Psi^{s}_{i}$ and $\Psi^{s'}_{j}$ states plotted against $\xi_{d}=\varepsilon_{d}+U_{d}/2$. Results are obtained for $t_{m}/U_{d}=0.3$, assuming a finite overlap between the edge states $\epsilon_m=0.3U_d$.
 } 
	\label{fig.amp_epsm}
\end{figure}

\begin{figure}
	\includegraphics[width=\wymiarm]{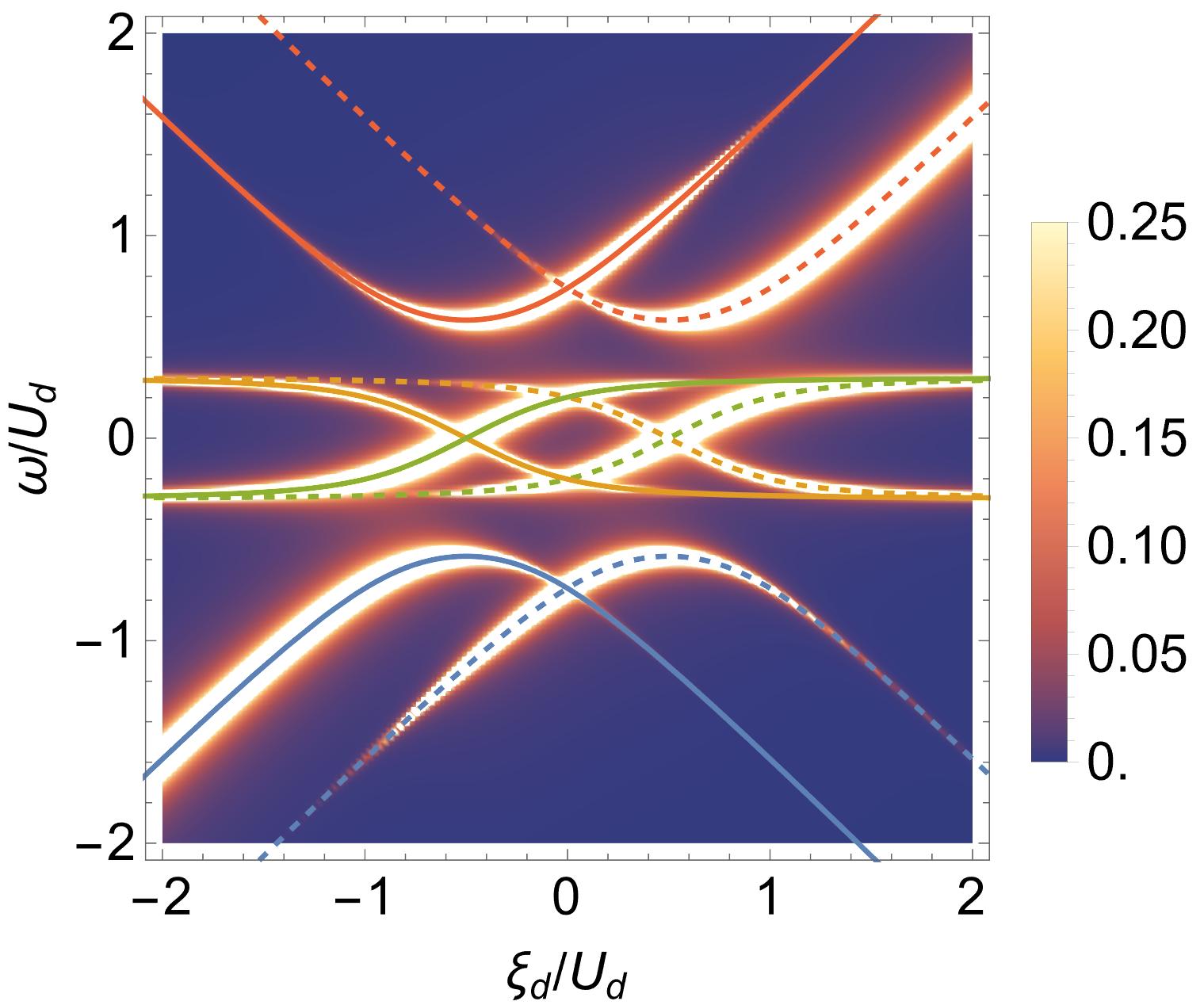}
	\caption{Variation of the quasiparticle spectrum 
$\rho_{d\downarrow}(\omega)$ with respect to $\xi_{d}=\varepsilon_d+U_d/2$ obtained for $t_m=0.25U_d$ and $\epsilon_m=0.3U_d$. The position of poles related to transitions between $\Psi_{1}$ and $\Psi_2$ states is marked with dashed lines, whereas transitions between $\Psi_{3}$ and $\Psi_4$ are marked with solid ones.
} 
	\label{dos2}
\end{figure}

For $\epsilon_m \neq 0$ we obtain nondegenerate eigenfunctions, characterized by 8 quasiparticle excitation energies. The transitions from each \(|\Psi_{i}\rangle\) to \(|\Psi_{j}\rangle\) are always accompanied by the corresponding transitions from \(|\Psi_{j}\rangle\) to \(|\Psi_{i}\rangle\) (with interchanged upper indexes), contributing to the quasiparticle energy. For instance, the transition \(|\Psi^{-}_{1}\rangle \rightarrow |\Psi^{-}_{2}\rangle\) contributes to the same pole as \(|\Psi^{+}_{2}\rangle \rightarrow |\Psi^{+}_{1}\rangle\). Although in general $\rho_{\downarrow}(\omega)$ is characterized by 8 quasiparticle energies, in practice all of them are visible only in close vicinity of the half-filling, $\xi_{d} \simeq 0$. Outside of this region, some amplitudes become negligible and the spectrum of $\downarrow$-spin electrons is represented by four quasiparticles (Figs. \ref{fig.amp_epsm} and \ref{dos2}). Far away from the half-filling, $|\xi_{d}| \gg U_d$, one pair disappears as well, and the spectrum simplifies to the standard single \JB{quasiparticle}.

\begin{figure}
	\includegraphics[width=\wymiarm]{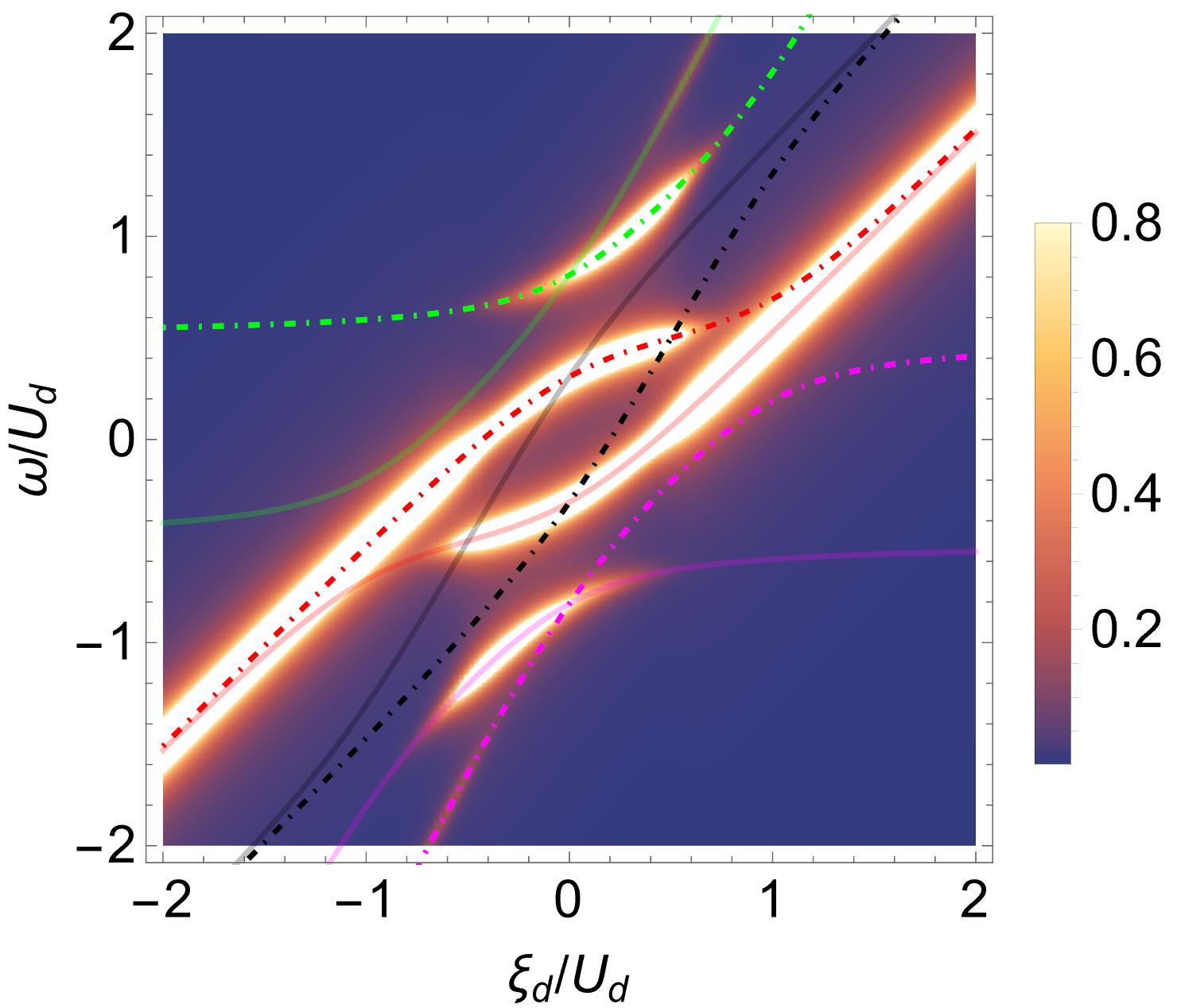}
	\caption{Variation of the quasiparticle spectrum 
 $\rho_{d\uparrow}(\omega)$ with respect to $\xi_{d}=\varepsilon_d+U_d/2$ obtained for $t_m=0.25U_d$ and $\epsilon_m=0.3U_g$. Position of poles related to transitions between $\Psi_{2}$ and $\Psi_3$ states are marked with dashed lines whereas transitions between  $\Psi_{1}$ and $\Psi_4$ are marked with solid faded ones.} 
	\label{dos2b}
\end{figure}

In Figure \ref{dos2} we plot the density of states $\rho_{\downarrow}(\omega)$ for nonvanishing $\epsilon_{m}$, which resembles the bowtie shapes obtained earlier \cite{Prada2017} from the mean-field approximation. Under specific conditions, $\xi_d=\pm\frac{U_d}{2}$, 
we observe a crossing of the Majorana features, which otherwise are split into bonding/antibonding energies.
At half-filling, the quasiparticle energies related to transitions $|\Psi^{s}_1 \rangle\leftrightarrow |\Psi^{s'}_2 \rangle$ are identical to the quasiparticle energies for transitions $|\Psi^{s}_3 \rangle \leftrightarrow |\Psi^{s'}_4 \rangle$, i.e., $E^{s}_{p}=E^{s}_{q}$. Consequently, the trivial and topological features are represented by four peaks at $\frac{1}{2}(\pm E_{p}^{-} \pm E_{p}^{+})$ and $\frac{1}{2}(\pm E_{q}^{-} \pm E_{q}^{+})$.

Nonvanishing $\epsilon_{m}$ also modifies the spectrum of electrons that are not directly coupled to MZM. Figure \ref{dos2b} shows that when a nonzero $\epsilon_m$ is introduced, the spectrum of $\uparrow$ electrons generally exhibits an 8-peak structure (marked by different color lines). However, four of these peaks have small amplitude across the entire energy range, while 2 additional peaks display significant amplitude only near half-filling. As a result, close to half-filling, we observe 4 well-pronounced peaks, whereas far from half-filling, only 2 peaks remain.

\section{Summary and outlook}

We have studied the spectrum of the single quantum dot coupled to the boundary
modes of the topological superconductor. From \JB{the} exact solution of this setup, 
we inferred the energies and spectral weights of the leaking \JB{Majorana} mode(s) coexisting with the conventional (nontopological) quasiparticles.

For \JB{the} non-correlated case, the trivial quasiparticles exist at energies $\pm \sqrt{\varepsilon_d^2 + 4t_m^2}$. In this scenario, the optimal amplitude of the zero-energy mode occurs for $\varepsilon_d = 0$. Under such circumstances, the Majorana mode acquires \JB{half} of the total spectral weight, and the trivial quasiparticles equally share the remaining amount. The spectral function of QD in this case is represented by a three-peak structure $\rho_{\downarrow}(\omega)=0.5\delta(\omega)+
0.25\delta(\omega-2t_m)+0.25\delta(\omega+2t_m)$. Away from half-filling, one of the trivial quasiparticles gradually \JB{absorbs} more and more spectral weight, at the expense of both the other \JB{conventional} quasiparticle and the zero-energy mode. 

In \JB{the} presence of the Coulomb repulsion, a leakage of the zero-energy mode is most efficient when the zero mode coincides either with the energy level $\varepsilon_d = 0$ or with the Coulomb satellite $\varepsilon_d +U_d=0$ (i.e. $\xi_d=\pm \frac{U_d}{2}$). 
One should note \JB{that}, at such points, the trivial quasiparticles are formed at $\pm 2t_m$, provided that the Majorana modes do not overlap with one another.
Away from these points, the spectrum of $\downarrow$-spin electrons consists of four trivial quasiparticles coexisting with the zero-energy mode (Fig.\ref{fig5pole}).
We have demonstrated that they could be experimentally detected by spin-polarized Andreev spectroscopy, Fig.\ \ref{fig.SESAR}.
The spin-$\uparrow$ sector also consists of four
quasiparticle branches, but all of them refer to the nontopological states.

We also investigated the quantum dot spectrum for the case of a short topological superconductor, where the Majorana modes overlap \JB{with} one another. In such \JB{a} situation, the boundary modes transmitted onto the correlated quantum dot form
two sets of bonding/antibonding states separated from the remaining four trivial quasiparticle branches, Fig.\ \ref{dos2}. Again, the optimal spectral weight of 
the topological quasiparticles coincides with 
$\xi_d=\pm \frac{U_d}{2}$. Near these special points, the Majorana modes cross each other, forming a bowtie shape.

Our analytical study extends the previous results \cite{Ricco-2019} obtained within the Hubbard-I approximation. \JB{The} expressions obtained here could be a useful starting point for further considerations of the many-body effects arising from the coupling of QD-MBS to mobile electrons of the external lead(s). They would also be helpful for investigating far-from-equilibrium effects, which can be induced by imposing quantum quench and/or periodic driving.


\appendix*
\section{Influence of Zeeman field}

In \JB{the} presence of the external magnetic field, $h$, the Hamiltonian of QD
takes the form
\begin{eqnarray}
\hat{H}_{QD}=(\varepsilon_{d}+h) \hat{d}_{\uparrow}^{\dagger}\hat{d}_{\uparrow}+(\varepsilon_{d}-h) \hat{d}_{\downarrow}^{\dagger}\hat{d}_{\downarrow}+U_d \hat{n}_{\uparrow}\hat{n}_{\downarrow}, 
\end{eqnarray}
where $\epsilon_{\downarrow} =\epsilon_{d}-h$, $\epsilon_{\uparrow} =\epsilon_{d}+h$ yield the eigenvalues (\ref{eq10}) 
\begin{eqnarray}
   E_{1}^{\pm}&=&\frac{1}{2}\left[ \epsilon_{\downarrow} \pm \sqrt{(\epsilon_{\downarrow}+\epsilon_{m})^2+4t_m^2} \right]
    \label{eqap1} \\ 
     E_{2}^{\pm}&=&\frac{1}{2}\left[ \epsilon_{\downarrow} \pm \sqrt{(\epsilon_{\downarrow}-\epsilon_{m})^2+4t_m^2} \right] 
     \label{eqap2} \\ 
     E_{3}^{\pm}&=&\!\frac{1}{2}\!\left[ \epsilon_{\downarrow}\!+\!2\epsilon_{\uparrow}\!+\!U_d\!\pm\! \sqrt{(\epsilon_{\downarrow}\!-\!\epsilon_{m}\!+\!U_{d})^2\!+\!4t_m^2} \right]\!
     \label{eqap3} \\ 
     E_{4}^{\pm}&=&\!\frac{1}{2}\!\left[ \epsilon_{\downarrow}\!+\!2\epsilon_{\uparrow}\!+\!U_d\!\pm\! \sqrt{(\epsilon_{\downarrow}\!+\!\epsilon_{m}\!+\!U_{d})^2\!+\!4t_m^2} \right].
     \label{eqap4}
\end{eqnarray}
Coefficients $u^{s}_i$ and $v_{i}^{s}$ remain the same upon substituting $\epsilon_{d}$ by $\epsilon_{\downarrow}$.
\begin{figure}
	\includegraphics[width=\wymiarm]{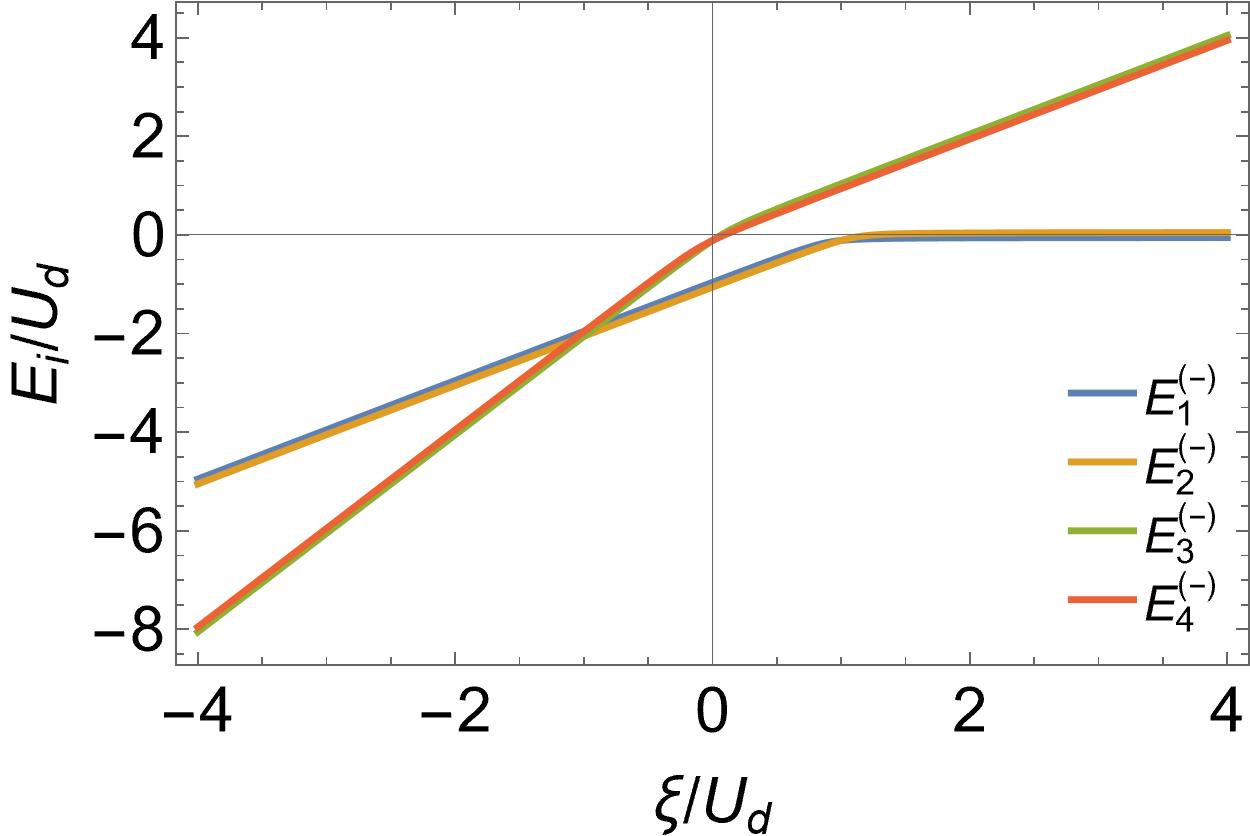}
	\caption{Dependence of the eigenenergies $E_{i}^{-}$ on the energy level $\varepsilon_{d}$ of QD in Zeeman field $h=0.5U_d$ obtained for weak overlap $\epsilon_{m}=0.1U_d$ .} 
	\label{fig.zemane}
\end{figure}
\begin{figure}[h]
	\includegraphics[width=\wymiarm]{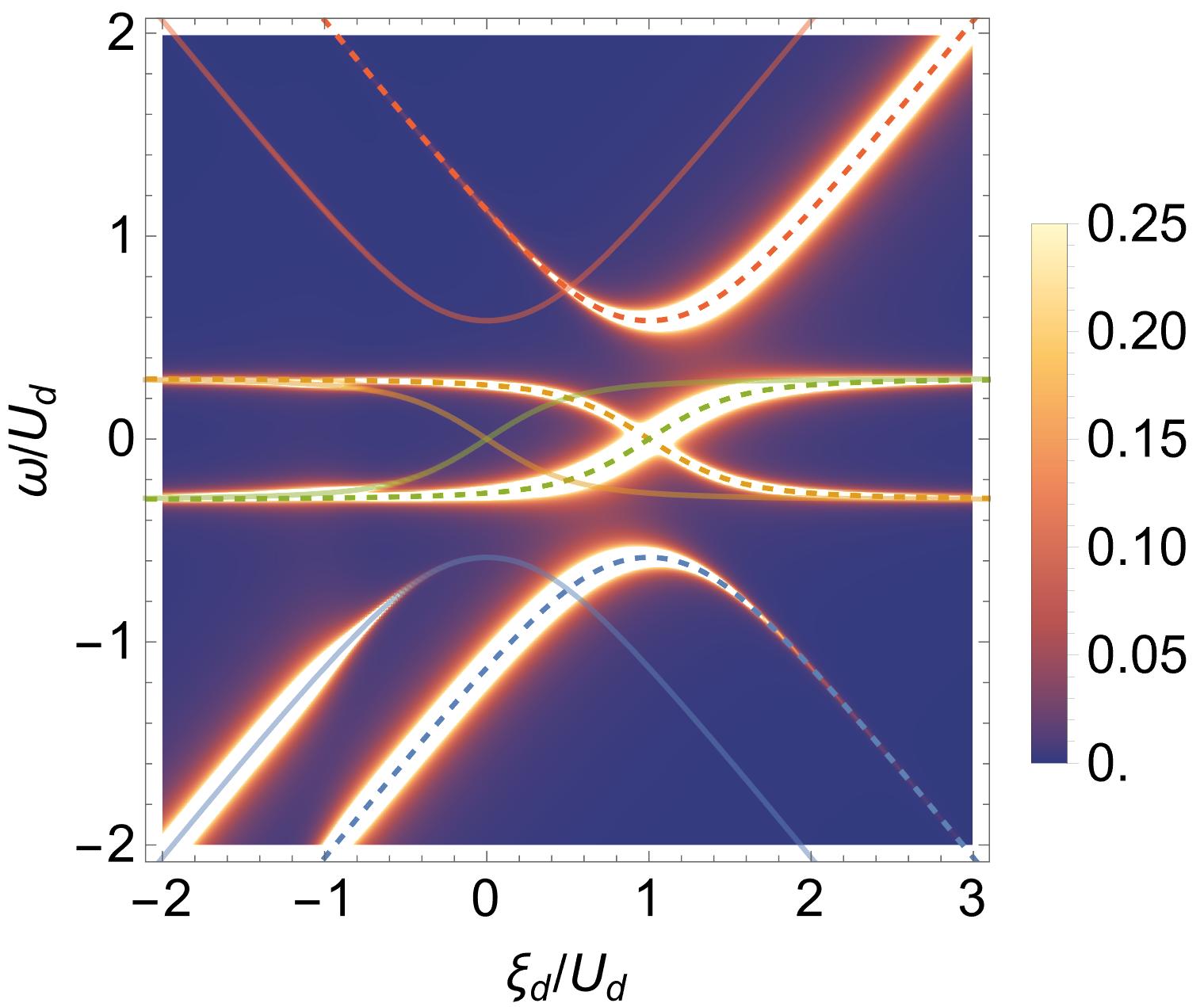}
	\caption{Variation of the quasiparticle spectrum 
 $\rho_{d\downarrow}(\omega)$ with respect to $\xi_{d}=\varepsilon_d+U_d/2$ obtained for the same set of parameters as in Fig. \ref{dos2} and Zeeman field $h=0.5U_d$. The position of poles related to transitions between $\Psi_{1}$ and $\Psi_2$ states is marked with dashed lines, whereas transitions between  $\Psi_{3}$ and $\Psi_4$ are marked with faded solid ones.} 
	\label{dos2_z1}
\end{figure}
A magnetic field parallel to $\downarrow$ spin causes lowering of energies $E_{1}$ and $E_{2}$. Conversely, $E_{3,4}$ have higher energies. This shifts the transition point from \JB{the} half-filling condition.
%
%
The Zeeman field shifts all quasiparticle peaks in both spin sectors, which can be seen in Figure \ref{dos2_z1}. Additionally, we notice that the magnetic field reduces the amplitudes of transitions between $\Psi_{3,4}$.

\section*{Data availability}
The datasets generated and analyzed during the current study are available
from the repository \cite{zenodo}.
\section*{Acknowledgements}
This research project has been supported by National Science Centre (Poland) through the grant 
no.\ 2022/04/Y/ST3/00061.
\clearpage
\bibliography{bibliography}

\end{document}